\documentclass[journal]{IEEEtran}
\usepackage{amsmath,amsfonts}
\usepackage{graphicx}
\usepackage[caption=false,font=normalsize,labelfont=sf,textfont=sf]{subfig}
\usepackage{amssymb}
\usepackage{algorithm}
\usepackage{algpseudocode}
\usepackage{acronym}
\usepackage{hyperref}
\usepackage{color}
\usepackage{tikz}
\usetikzlibrary{positioning, shapes.geometric, arrows.meta,
  decorations.pathreplacing, calc, patterns, backgrounds, shapes}
\usepackage{xcolor}
\usepackage{graphicx}
\usepackage{booktabs}
\usepackage{multirow}
\usepackage{tabularx}
\usepackage{dsfont}

\acrodef{ABR}{Adaptive Bitrate}
\acrodef{ACPF}{Adaptive Cheapest Pipe First}
\acrodef{APR}{Average Playback Resolution}
\acrodef{ATSSS}{Access Traffic Steering, Switching, and Splitting}
\acrodef{ATSSS-HL}{ATSSS High Layer}
\acrodef{ATSSS-LL}{ATSSS Low Layer}
\acrodef{BPF}{Best Path First}
\acrodef{CPF}{Cheapest Path First}
\acrodef{CWND}{Congestion Window}
\acrodef{DAPS}{Delay-Aware Packet Scheduling}
\acrodef{DARA}{Deep Adaptive Rate Allocation}
\acrodef{DQN}{Deep Q-Network}
\acrodef{DRL}{Deep Reinforcement Learning}
\acrodef{ECF}{Earliest Completion First}
\acrodef{E2E}{End-to-End}
\acrodef{FTP}{File Transfer Protocol}
\acrodef{HoL}{Head-of-Line}
\acrodef{MAE}{Mean Absolute Error}
\acrodef{MAMS}{Mobility-Aware Multipath Scheduler}
\acrodef{ML}{Machine Learning}
\acrodef{minRTT}{Minimum Round-Trip Time}
\acrodef{MP-DCCP}{Multipath Datagram Congestion Control Protocol}
\acrodef{MP-QUIC}{Multipath QUIC}
\acrodef{MP-TCP}{Multipath TCP}
\acrodef{MSE}{Mean Squared Error}
\acrodef{NAS}{Neural Architecture Search}
\acrodef{NRMSE}{Normalised Root Mean Square Error}
\acrodef{OFO}{Out-of-Order}
\acrodef{OTIAS}{Out-of-Order Transmission for In-order Arrival Scheduler}
\acrodef{PPO}{Proximal Policy Optimisation}
\acrodef{QoS}{Quality of Service}
\acrodef{RMSE}{Root Mean Square Error}
\acrodef{RLMR}{Reinforcement Learning Based Multipath Routing}
\acrodef{RR}{Round Robin}
\acrodef{RTT}{Round-Trip Time}
\acrodef{SDN}{Software-Defined Networks}
\acrodef{SRTT}{Smoothed Round-Trip Time}
\acrodef{SWND}{Sending Window}
\acrodef{UE}{User Equipment}
\acrodef{BLEST}{BLocking ESTimation}
\acrodef{QoE}{Quality of Experience}
\acrodef{RL}{Reinforcement Learning}
\acrodef{LSTM}{Long Short-Term Memory}
\acrodef{VoD}{Video on Demand}
\acrodef{MDP}{Markov Decision Process}

\begin{document}
\title{Deep Adaptive Rate Allocation in Volatile
  Heterogeneous Wireless Networks}

\author{Gregorio Maglione, Veselin Rakocevic, Markus Amend,
  and Touraj Soleymani
\thanks{G.~Maglione, V.~Rakocevic, and T.~Soleymani are with the Department of Engineering, 
  City St~George's, University of London,
  London EC1V~0HB, United Kingdom
  (e-mails: {\tt\small {gregorio.maglione, veselin.rakocevic.1,
      touraj.soleymani}@citystgeorges.ac.uk}).
  M.~Amend is with Deutsche Telekom, Darmstadt, Germany
  (e-mail: {\tt\small markus.amend@telekom.de}).}}

\maketitle

\begin{abstract}
Modern multi-access 5G+ networks provide mobile terminals with
additional capacity, improving network stability and performance.
However, in highly mobile environments such as vehicular networks,
supporting multi-access connectivity remains challenging.
The rapid fluctuations of wireless link quality often outpace the
responsiveness of existing multipath schedulers and transport-layer
protocols. This paper addresses this challenge by integrating
Transformer-based path state forecasting with a new multipath
splitting scheduler called Deep Adaptive Rate Allocation (DARA).
The proposed scheduler employs a deep reinforcement learning engine
to dynamically compute optimal congestion window fractions on
available paths, determining data allocation among them.
A six-component normalised reward function with weight-mediated
conflict resolution drives a DQN policy that eliminates the
observation-reaction lag inherent in reactive schedulers.
Performance evaluation uses a Mininet-based Multipath Datagram
Congestion Control Protocol testbed with traces from mobile users in
vehicular environments. Experimental results demonstrate that DARA
achieves better file transfer time reductions compared to
learning-based schedulers under moderate-volatility traces. For
buffered video streaming, resolution improvements are
maintained across all tested conditions. Under controlled burst
scenarios with sub-second buffer constraints, DARA achieves
substantial rebuffering improvements whilst state-of-the-art
schedulers exhibit near-continuous stalling.
\end{abstract}


\section{Introduction}

Modern multi-access 5G networks provide mobile terminals with
additional capacity, improving overall network stability and
performance.  However, the reduction in cellular service area due to
the higher frequency ranges of 5G and the significantly smaller cell
areas of heterogeneous networks results in more frequent horizontal
handovers over the same physical distance, causing unstable
transmissions characterised by frequent bursts, volatile \ac{RTT}, and
periods of low throughput.  The rapid fluctuations of wireless link
quality in such highly mobile environments often outpace the
responsiveness of existing multipath schedulers and multipath
transport-layer protocols.  Paradigms such as 5G's
\ac{ATSSS}~\cite{rel16} combine 3GPP and non-3GPP paths (e.g., 5G
and Wi-Fi) to support unified multipath operation.  In particular,
ATSSS defines two complementary mechanisms.  While \ac{ATSSS-LL}, as
a link-layer mechanism, provides basic traffic steering using network
or local policies, the core enabler is \ac{ATSSS-HL}, which, as a
transport-layer mechanism, provides traffic splitting using congestion
states as well as path availability, latency, and load.  A multipath
protocol can be used to implement \ac{ATSSS-HL} functionality, with
scheduling, path estimation, and packet reordering, allowing packets
to be enumerated and sent on the relevant path despite \ac{OFO}
delivery.  Common multipath protocols in this context include:
\ac{MP-TCP}~\cite{rfc8684},
\ac{MP-QUIC}~\cite{ietf-quic-multipath-17}, and
\ac{MP-DCCP}~\cite{rfc9897}. Nevertheless, all
these conventional strategies, which primarily rely on reactive path
switching between coexisting communication infrastructures, are often
ineffective in high mobility environments.  Consequently, ensuring
robust and efficient communication over volatile heterogeneous
networks demands a rethinking of rate allocation strategies, moving
beyond static or rule-based scheduling towards adaptive, data-driven
approaches.

While existing learning-based approaches incorporate prediction
mechanisms, these solutions still face limitations in capturing rapid
sub-second traffic variations characteristic of high-mobility
scenarios.  In this paper, a novel framework referred to as \ac{DARA}
is developed to fully exploit burst capacity in volatile heterogeneous
networks.  \ac{DARA} consists of two modules: an offline module and an
online module.  The former employs a Transformer model for time-series
analysis of historical congestion control states, forecasting future
\ac{CWND} and \ac{SRTT} values at five horizons (100--500\,ms) per
path.  These predictions are integrated with current observations to
form an 18-dimensional state vector comprising absolute metrics,
normalised rate-of-change deltas, previous actions, and binary
congestion flags.  A \ac{DQN} operates on this state asynchronously to
solve a multi-objective optimisation problem formulated to balance
throughput maximisation, delay minimisation, and path utilisation
fairness.  The output is a per-path \ac{CWND} utilisation fraction,
enabling fine-grained per-path rate control that proactively exploits
predicted burst opportunities and avoids predicted congestion states
before they materialise, thereby circumventing the
observation-reaction lag that fundamentally limits reactive schedulers
in high-mobility scenarios.

\subsection{Related Work}

Multipath schedulers~\cite{9645537} aim to exploit the availability of
multiple concurrent paths by routing traffic according to different
priorities such as throughput maximisation, transmission cost
reduction, or latency minimisation. In throughput-oriented designs,
load balancing plays a central role, as in the case of \ac{RLMR} in
\ac{SDN}~\cite{10.1155/2022/5124960}, or path aggregation schemes in
last-mile networking. Cost-oriented scheduling is often motivated by
operator-side requirements, exemplified by the \ac{CPF}
scheduler~\cite{amend2019framework}. In heterogeneous wireless
networks, however, latency reduction is typically the most critical
objective due to pronounced path asymmetry. To mitigate such latency
effects, a number of scheduling approaches have been proposed:
\ac{minRTT} scheduling~\cite{rfc8684}, which sends packets over the
path with the lowest \ac{RTT}; \ac{OTIAS}~\cite{6844729} and
\ac{DAPS}~\cite{6883488}, which deliberately transmit packets
out-of-order to achieve in-order arrival at the receiver;
\ac{ECF}~\cite{10.1145/3143361.3143376}, which minimises idle time on
the fast path; and \ac{BLEST}~\cite{7497206}, which predicts and
mitigates \ac{HoL} blocking.

Dong \textit{et al.}~\cite{8636963} conclude in a survey that
\ac{BLEST} generally outperforms other \ac{MP-TCP} schedulers,
although in non-shared bottleneck conditions most schedulers converge
to similar performance. This highlights that path-estimation
schedulers are inherently suboptimal in volatile conditions, where
frequent path fluctuations aggravate bottlenecks, increase packet
reordering, and exacerbate \ac{HoL} blocking. To overcome these
limitations, Diakhate \textit{et al.}~\cite{10.1145/3568562.3568655}
compare rule-based schedulers with intelligent schedulers, concluding
that adaptivity is essential in mobility scenarios. This is because
the inherent path asymmetry introduces packet scrambling, \ac{OFO}
delivery, and \ac{HoL} blocking~\cite{7497206}; the adverse
interactions between multipath tunnelling and nested congestion
control mechanisms limit path aggregation~\cite{10.1145/3472305.3472316};
and the layered interactions between multipath tunnels and congestion
control mechanisms can even negate throughput
gains~\cite{fi16070233}. The resulting body of \ac{ML}-driven work
can be broadly categorised into: (i)~redundancy-based
solutions~\cite{10053647,10049750},
(ii)~cross-layer solutions~\cite{9142742,101145,10486871},
(iii)~estimation-based
solutions~\cite{10060683,8802063,8666496,ZENG2022109198,220108969,9110610},
and (iv)~multi-objective
solutions~\cite{9826510,8737649,10.1145/3649139}.

\paragraph{Redundancy-Based Solutions}
Redundancy-based scheduling maintains \ac{QoS} by predicting link
quality and sending redundant packets on secondary paths when outage
or degradation is predicted. Xing \textit{et al.}~\cite{10049750}
optimise the redundancy ratio using a multi-armed bandit framework,
whilst Han \textit{et al.}~\cite{10053647} use a \ac{DRL} agent to
optimise goodput and bandwidth utilisation during handoffs. Both
approaches impose significant resource costs that scale poorly with
the number of users.

\paragraph{Cross-Layer Solutions}
Cross-layer solutions exchange control information between OSI layers.
Lv \textit{et al.}~\cite{101145} combine server transport-layer
scheduling with client application-level \ac{ABR} streaming, and
Yang \textit{et al.}~\cite{10486871} introduce \ac{MAMS}, which uses
forecasting to find congestion and buffer utilisation based on
historical path conditions and wireless uplink/downlink data. Both
show that \ac{QoS} improves substantially when path parameters are
known in real time, but require protocol modifications precluding
universal adoption.

\paragraph{Prediction-Based Solutions}
Prediction-based scheduling estimates path conditions without direct
receiver-state feedback. Luo \textit{et al.}~\cite{8666496} propose a
heuristic \ac{SRTT}-bandwidth scheduler; Nguyen
\textit{et al.}~\cite{10060683} use Q-learning-based path selection;
and Rosell\'{o}~\cite{8802063} applies \ac{DRL} for higher-dimensional
states. These greedy methods perform poorly in rapidly changing
environments. Zeng \textit{et al.}~\cite{ZENG2022109198} assign data
based on estimated transfer completion times. Wu
\textit{et al.}~\cite{9110610} develop Peekaboo and subsequently the
FALCON scheduler~\cite{220108969}, both combining offline
coarse-grained and online fine-grained models using transport-layer
parameters.

\paragraph{Multi-Objective Solutions}
Multi-objective optimisation balances trade-offs where single-objective
schedulers fail. Zhang \textit{et al.}~\cite{8737649} define a
multi-objective reward balancing delay, throughput, and loss solved
via \ac{DRL}. Han \textit{et al.}~\cite{10.1145/3649139} use
Multi-Agent \ac{DRL} with per-path neural networks to minimise
bufferbloat. Dong \textit{et al.}~\cite{9826510} propose MPTCP-RL,
optimising energy efficiency, throughput, delay, and jitter using a
\ac{PPO} model.

\subsection{Main Contributions and Outline of the Paper}

In this paper, DARA is proposed: a predictive multipath scheduler
combining Transformer-based congestion forecasting with DQN-based rate
control for volatile heterogeneous networks. The main contributions
are:

\begin{itemize}
\item \textbf{Predictive rate control architecture:} Combines
  multi-horizon (100--500\,ms) Transformer predictions at 100\,ms
  aggregated resolution with continuous per-path \ac{CWND} fraction
  adjustment, eliminating the observation-reaction lag inherent in
  burst exploitation.

\item \textbf{Multi-objective DQN formulation with normalised rewards
  and conflict resolution:} A six-component reward function in which
  all components are verified to occupy $\mathcal{O}([-1,1])$ ranges,
  ensuring that optimised weights directly encode relative objective
  importance. Weight-mediated conflict resolution is analysed for
  three principal conflict scenarios, and a systematic weight search
  over 50 iterations identifies the deployed configuration.

\item \textbf{Protocol-agnostic validation framework:}
  Production-grade \ac{MP-DCCP} testbed with custom kernel-userspace
  interface and trace-driven Mahi-Mahi emulation enabling ATSSS-HL
  validation across diverse traffic (FTP, YouTube, live streaming)
  under real mobility patterns.

\item \textbf{Comprehensive experimental analysis with ablation
  validation:} Design decisions substantiated through predictor
  architecture selection, Transformer depth analysis, action-space
  granularity sensitivity, and full component ablation with
  statistical validation (Welch's $t$-test, Mann-Whitney $U$,
  Cohen's $d$, bootstrap confidence intervals), followed by
  systematic evaluation against 8~schedulers across 5~cellular
  traces and synthetic burst scenarios.
\end{itemize}

The rest of the paper is organised as follows.
Section~\ref{sec:DARA} delineates the proposed framework, detailing
the Transformer architecture, multi-objective reward formulation, and
DQN-based scheduling algorithm.
Section~\ref{sec:methodology} describes the Mininet \ac{MP-DCCP}
testing methodology. \ac{MP-DCCP} is chosen because its unreliable
delivery avoids nested reliability conflicts when encapsulating
TCP/QUIC, eliminates mandatory \ac{HoL} blocking unlike MPTCP/MP-QUIC,
and provides kernel-level integration with mature Linux support.
Section~\ref{sec:results} presents experimental results, opening with
ablation studies that substantiate each architectural decision,
followed by controlled burst and real-world trace-driven performance
comparisons against state-of-the-art schedulers.

\section{Deep Adaptive Rate Allocation Framework}
\label{sec:DARA}

This section presents the \ac{DARA} framework. It begins by
establishing DARA's core novelty of predictive rate control through
\ac{CWND} fraction adjustment, then details the Transformer-based
prediction model, the \ac{DQN} policy network, and the integrated
training and execution algorithm.

\subsection{Overview and Novelties of DARA}

DARA introduces a fine-grained predictive rate control mechanism based
on a Transformer model. Given that channel-quality-induced bursts due
to mobility-driven fluctuation occur over short time intervals,
frequently in the order of 100\,ms, reactive schedulers fail to fully
exploit path capacity during bursts and fail to reduce transmission
rates once bursts cease, leading to capacity waste and increased
\ac{OFO} packets.

In this deployment, each subflow runs independent BBR-like congestion
control maintaining a separate \ac{CWND}. The scheduler distributes
connection-level data across available paths whilst respecting each
path's \ac{CWND} constraint. Schedulers that select paths based on
instantaneous metrics create a fundamental coupling problem: by the
time the scheduler observes \ac{CWND} changes, transmission
opportunities have already been missed or paths have been
oversubscribed.

To address this, a \ac{DQN} policy network outputs discrete \ac{CWND}
fraction actions, effectively rate-limiting each path by restricting
its usable congestion window to $\phi_p \times \text{CWND}_p$. This
multiplicative control directly constrains the scheduler's
transmission decisions: when path~$p$ has $\text{CWND}_p = 100$
segments and \ac{DARA} sets $\phi_p = 0.3$, the scheduler treats the
effective capacity as only 30~segments, preventing oversubscription
before congestion materialises. As illustrated in
Figure~\ref{fig:framework}, when \ac{DARA} predicts Path~1
degradation (declining \ac{CWND}, rising \ac{SRTT}), it proactively
limits the path to a reduced capacity ($\phi_1 = 0.3$) whilst
allowing stable Path~2 full utilisation ($\phi_2 = 1.0$), resulting
in near-sequential packet arrival at the receiver and minimal
\ac{HoL} blocking.

\begin{figure}[t]
\centering
\includegraphics[width=\linewidth]{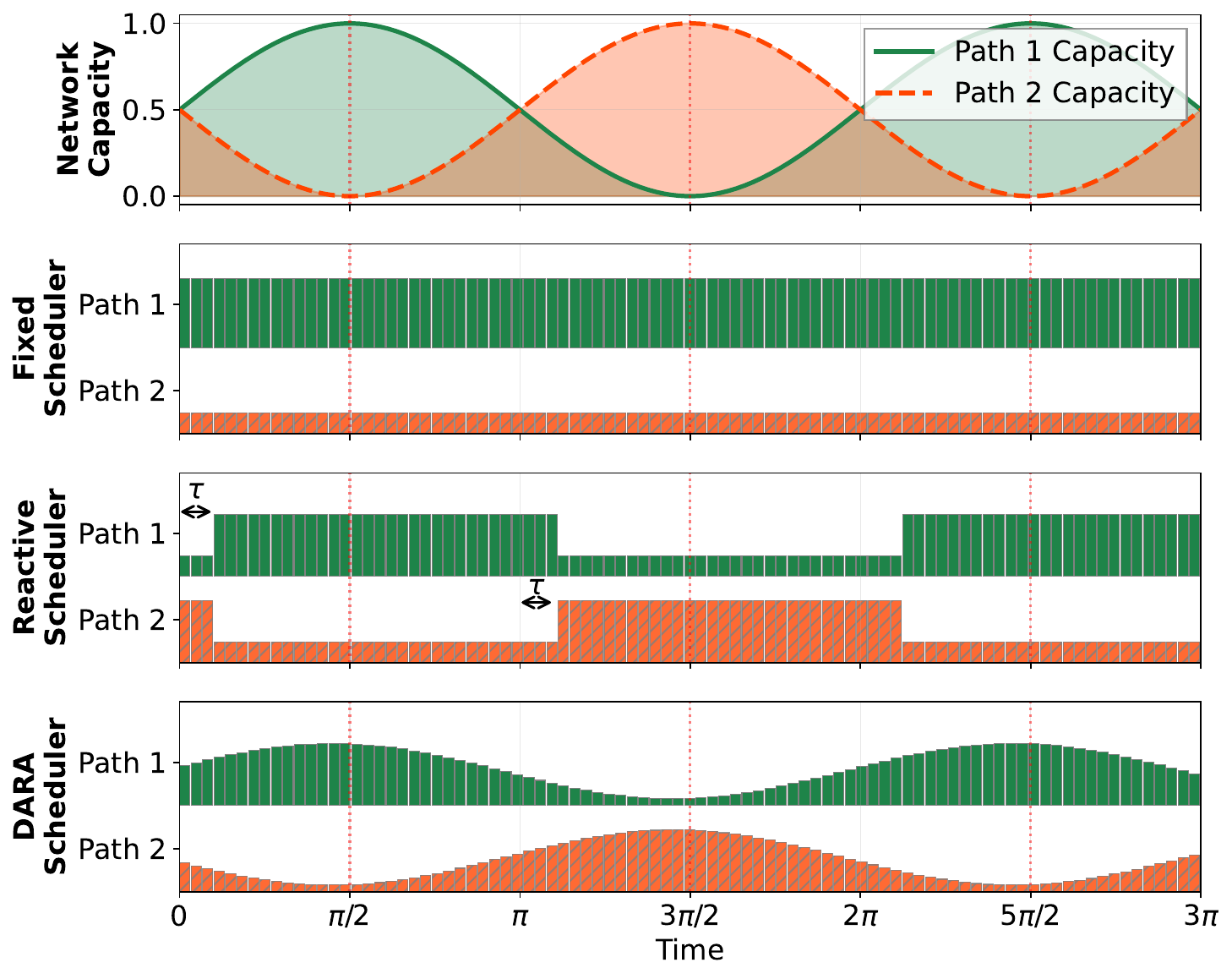}
\caption{Scheduler response to antiphase path capacity variations.
  Fixed-rule schedulers maintain fixed allocation regardless of path
  capacity. Reactive schedulers exhibit observation-reaction
  lag~$\tau$. DARA eliminates this lag through its predictive
  horizon, preemptively adjusting CWND fractions to match anticipated
  capacity variations.}
\label{fig:Scheduler-Comparison}
\end{figure}

The key advantage over reactive approaches is predictive
\ac{CWND} fraction adjustment: whilst congestion control algorithms
adjust \ac{CWND} reactively over multiple RTT timescales, \ac{DARA}
adjusts the scheduler's utilisation of that \ac{CWND} predictively,
enabling sub-RTT adaptation. As shown in
Figure~\ref{fig:Scheduler-Comparison}, static schedulers maintain
fixed allocation regardless of \ac{CWND} evolution, and reactive
schedulers (e.g., \ac{CPF}) exhibit observation-reaction
lag~$\tau$. \ac{DARA} eliminates this lag through its predictive
horizon.

While existing advanced schedulers incorporate forward-looking
elements, such as offline meta-learning in
FALCON~\cite{220108969} or cross-layer mobility awareness in
\ac{MAMS}~\cite{10486871}, DARA uniquely combines
high-temporal-resolution prediction with continuous per-path rate
control through \ac{CWND} fraction adjustment. This distinguishes
\ac{DARA} from: (1)~discrete-action \ac{DRL} schedulers that select
paths binarily~\cite{8802063}; (2)~binary decision frameworks lacking
continuous rate adjustment~\cite{220108969}; (3)~specialised
optimisers with narrower objectives~\cite{9826510,9142742}; and
(4)~cross-layer approaches requiring protocol
modifications~\cite{101145,10486871}.
Figure~\ref{fig:framework} illustrates the DARA architecture.

\definecolor{dara_color}{RGB}{227, 114, 34}
\definecolor{path1_color}{RGB}{31, 119, 180}
\definecolor{path2_color}{RGB}{214, 39, 40}
\definecolor{buffer_color}{RGB}{144, 238, 144}
\begin{figure}[h!]
\centering
\begin{tikzpicture}[
    scale=0.7, transform shape,
    node distance=0.75cm,
    block/.style={
        rectangle, draw, thick, fill=gray!10,
        text width=2.8cm, text centered, rounded corners=2pt,
        minimum height=0.85cm, font=\sffamily\small
    },
    dara_block/.style={
        rectangle, draw=dara_color, very thick, fill=dara_color!15,
        text width=7cm, minimum height=2cm, rounded corners=3pt
    },
    ml_component/.style={
        rectangle, draw, thick, fill=blue!10,
        text width=1.5cm, text centered, rounded corners=2pt,
        minimum height=0.5cm, font=\sffamily\scriptsize
    },
    buffer_visual/.style={
        rectangle, draw, thick,
        minimum width=2.6cm, minimum height=0.65cm, rounded corners=2pt
    },
    path_block/.style={
        rectangle, draw, thick, fill=gray!10,
        text width=2.6cm, text centered, rounded corners=2pt,
        minimum height=0.75cm, font=\sffamily\small
    },
    packet/.style={
        rectangle, draw, thick,
        minimum width=0.22cm, minimum height=0.28cm, rounded corners=0.5pt,
        inner sep=0pt
    },
    arrow/.style={
        -{Stealth[length=2.5mm, width=1.8mm]}, thick
    },
    data_arrow/.style={
        arrow, black, line width=1pt
    },
    control_arrow/.style={
        arrow, dara_color, dashed, line width=0.8pt
    }
]
\node[block] (app) at (0, 0) {Sender\\Application};
\node[dara_block, below=0.65cm of app] (dara) {};
\node[font=\sffamily\bfseries\small] at (dara.north) [yshift=-0.2cm] {DARA Controller};
\node[ml_component] (transformer) at ($(dara.center) + (-1.2, 0.1)$) {Transformer\\Predictions};
\node[ml_component] (dqn) at ($(dara.center) + (1, 0.1)$) {DQN\\Policy};
\node[font=\sffamily\scriptsize, text width=5.1cm, align=center] at ($(dara.center) + (0, -0.55)$)
    {Observes: CWND$_i$, SRTT$_i$, $\rho_i$ $\rightarrow$ Predicts 500\,ms $\rightarrow$ Action: $[\phi_1, \phi_2]$};
\draw[-{Stealth[length=2mm, width=1.5mm]}, blue!70, line width=0.8pt] (transformer.east) -- (dqn.west);
\node[block, below=0.8cm of dara] (buffer_label) {Send Buffer};
\node[buffer_visual, below=0.05cm of buffer_label, fill=gray!15] (buffer_vis) {};
\foreach \x in {0,1,...,9} {
    \node[packet, fill=gray!50] at ($(buffer_vis.west) + (0.22+\x*0.24, 0)$) {\scriptsize\x};
}
\node[block, below=0.7cm of buffer_vis] (scheduler) {Scheduler};
\coordinate (path1_start) at ($(scheduler.south)+(-2.5,-1.2)$);
\node[path_block, fill=path1_color!10] (sf1) at (path1_start) {Path 1 Subflow};
\node[buffer_visual, below=0.05cm of sf1, minimum height=0.65cm] (cwnd1) {};
\foreach \x in {0,1,2,3,4,5} {
    \draw[thick, path1_color!20, fill=path1_color!5] ($(cwnd1.west) + (0.25+\x*0.38, -0.15)$) rectangle +(0.26, 0.35);
}
\foreach \x in {0,1} {
    \draw[thick, path1_color!70, fill=path1_color!30] ($(cwnd1.west) + (0.25+\x*0.38, -0.15)$) rectangle +(0.26, 0.35);
}
\node[font=\sffamily\tiny] at ($(cwnd1.north)+(0,0.20)$) {CWND$_1$ $\times$ $\phi_1$=0.3 (rate-limited)};
\node[buffer_visual, below=0.05cm of cwnd1, fill=path1_color!15, minimum height=0.5cm] (if1) {};
\node[packet, fill=path1_color!60] at ($(if1.west) + (0.85, 0)$) {\scriptsize 0};
\node[packet, fill=path1_color!60] at ($(if1.west) + (1.70, 0)$) {\scriptsize 2};
\node[path_block, below=0.55cm of if1, fill=path1_color!20] (cc1) {DCCP/CC};
\coordinate (path2_start) at ($(scheduler.south)+(2.5,-1.2)$);
\node[path_block, fill=path2_color!10] (sf2) at (path2_start) {Path 2 Subflow};
\node[buffer_visual, below=0.05cm of sf2, minimum height=0.65cm] (cwnd2) {};
\foreach \x in {0,1,2,3,4,5} {
    \draw[thick, path2_color!70, fill=path2_color!30] ($(cwnd2.west) + (0.25+\x*0.38, -0.15)$) rectangle +(0.26, 0.35);
}
\node[font=\sffamily\tiny] at ($(cwnd2.north)+(0,0.20)$) {CWND$_2$ $\times$ $\phi_2$=1.0 (full capacity)};
\node[buffer_visual, below=0.05cm of cwnd2, fill=path2_color!15, minimum height=0.5cm] (if2) {};
\foreach \x/\label in {0/1,1/3,2/4,3/5,4/6,5/7,6/8,7/9} {
    \node[packet, fill=path2_color!60] at ($(if2.west) + (0.23+\x*0.29, 0)$) {\scriptsize\label};
}
\node[path_block, below=0.55cm of if2, fill=path2_color!20] (cc2) {DCCP/CC};
\coordinate (bottom_center) at ($(cc1.south)!0.5!(cc2.south)$);
\node[buffer_visual, below=1.2cm of bottom_center, fill=buffer_color!25] (recv_buffer) {};
\foreach \x in {0,1,...,9} {
    \node[packet, fill=buffer_color!70] at ($(recv_buffer.west) + (0.22+\x*0.24, 0)$) {\scriptsize\x};
}
\node[font=\sffamily\scriptsize, above=0.05cm of recv_buffer] {Reorder Buffer};
\node[block, below=0.55cm of recv_buffer] (receiver) {Receiver\\Application};
\draw[data_arrow] (app) -- (dara);
\draw[data_arrow] (dara) -- (buffer_label);
\draw[data_arrow] (buffer_vis) -- (scheduler);
\draw[data_arrow] (scheduler.south) -- ++(0,-0.2) -| (sf1.north);
\draw[data_arrow] (scheduler.south) -- ++(0,-0.2) -| (sf2.north);
\draw[data_arrow] (if1) -- (cc1);
\draw[data_arrow] (if2) -- (cc2);
\draw[data_arrow] (cc1.south) -- ++(0,-0.6) -| ($(recv_buffer.north west)+(0.3,0)$);
\draw[data_arrow] (cc2.south) -- ++(0,-0.6) -| ($(recv_buffer.north east)+(-0.3,0)$);
\draw[data_arrow] (recv_buffer) -- (receiver);
\draw[control_arrow] (cc1.west) -- ++(-0.8,0) |- ($(dara.west)+(0,-0.3)$);
\draw[control_arrow] (cc2.east) -- ++(0.8,0) |- ($(dara.east)+(0,-0.3)$);
\node[font=\sffamily\small, align=center, fill=white, inner sep=2pt, rounded corners=1pt]
    at ($(dara.south)!0.5!(buffer_label.north)$) {CWND fractions\\$[\phi_1,\phi_2]$};
\node[font=\sffamily\scriptsize, fill=white, inner sep=2pt, rounded corners=1pt]
    at ($(buffer_vis.south)+(0,-0.17)$) {Segments awaiting scheduling};
\node[font=\sffamily\scriptsize, text width=3.2cm, align=center, fill=white, inner sep=2pt, rounded corners=1pt]
    at ($(scheduler.south)+(0,-0.45)$)
    {Applies CWND fractions $[\phi_1, \phi_2]$};
\node[font=\sffamily\scriptsize, fill=white, inner sep=2pt, rounded corners=1pt]
    at ($(if1.south)+(0,-0.20)$) {Limited in-flight};
\node[font=\sffamily\scriptsize, fill=white, inner sep=2pt, rounded corners=1pt]
    at ($(if2.south)+(0,-0.20)$) {Full utilisation};
\node[font=\sffamily\scriptsize, text width=2cm, align=left, fill=white, fill opacity=0.85, text opacity=1, inner sep=2pt, rounded corners=1pt]
    at ($(cc1.west)+(-1.3,-0.0)$)
    {CWND$_1$, SRTT$_1$, In-flight$_1$, Loss$_1$, Queue$_1$};
\node[font=\sffamily\scriptsize, text width=2cm, align=left, fill=white, fill opacity=0.85, text opacity=1, inner sep=2pt, rounded corners=1pt]
    at ($(cc2.east)+(1.3,-0.0)$)
    {CWND$_2$, SRTT$_2$, In-flight$_2$, Loss$_2$, Queue$_2$};
\node[font=\sffamily\scriptsize, text width=2.2cm, align=left,
    fill=yellow!20, draw=orange!50, rounded corners=2pt,
    inner sep=3pt]
    at ($(cc1.west)+(-1.1,1.5)$)
    {\textbf{Predicted:}\\CWND$_1
$$
\downarrow$, SRTT$_1
$$
\uparrow$\\$\Rightarrow$ $\phi_1$=0.3\\(limit to 30\%)};
\node[font=\sffamily\scriptsize, text width=2cm, align=left,
    fill=green!15, draw=green!50, rounded corners=2pt,
    inner sep=3pt]
    at ($(cc2.east)+(1.1,1.5)$)
    {\textbf{Predicted:}\\CWND$_2$ stable\\$\Rightarrow$ $\phi_2$=1.0\\(full capacity)};
\node[font=\sffamily\scriptsize, text width=2.3cm, align=center,
    fill=buffer_color!40, draw=green!60, rounded corners=2pt,
    inner sep=3pt]
    at ($(recv_buffer.east)+(1.6,0)$)
    {\textbf{Result:}\\Predictive rate\\limiting $\rightarrow$\\sequential arrival\\$\rightarrow$ minimal\\HoL blocking};
\end{tikzpicture}
\caption{DARA's Predictive Rate Control Architecture. The Transformer
  predicts CWND and SRTT evolution for each path. The DQN policy uses
  these predictions to compute CWND fractions that rate-limit each
  path. In this example, DARA predicts Path~1 degradation and
  pre-emptively limits it to $\phi_1=0.3$, whilst allowing the stable
  path full utilisation at $\phi_2=1.0$, enabling sequential packet
  arrival and minimal head-of-line blocking.}
\label{fig:framework}
\end{figure}
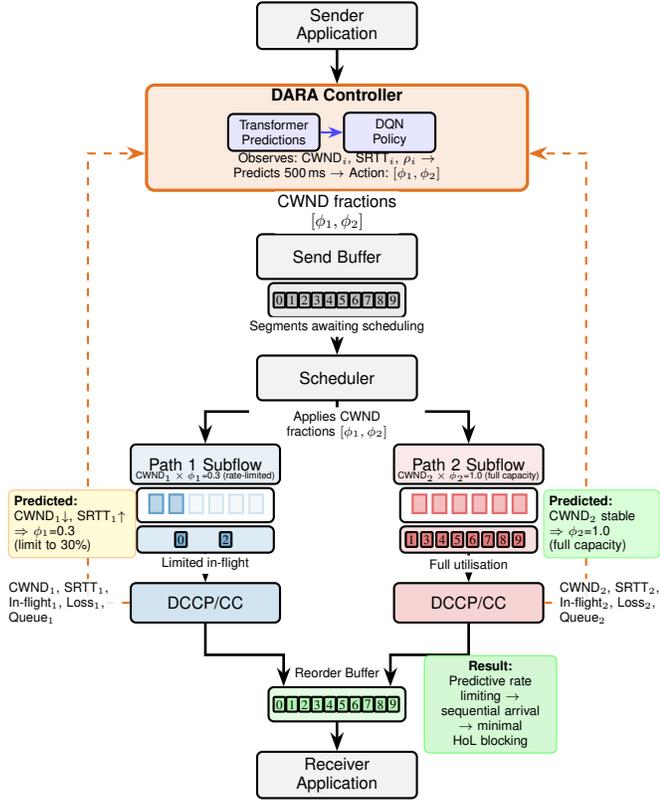

The inference pipeline completes in approximately 3\,ms. \ac{DARA}
operates on 100\,ms control cycles, computing optimal \ac{CWND}
fractions that the packet scheduler applies at per-packet granularity
between updates.

\subsection{Transformer-Based Prediction Model}
\label{sec:transformer}

To enable predictive scheduling, a Transformer model with multi-head
attention trained on congestion control telemetry forecasts \ac{CWND}
and \ac{SRTT} trajectories. The architecture processes sequences of
8~aggregated time steps (each representing 100\,ms bins) to predict
values at 5~future horizons (100, 200, 300, 400, 500\,ms) for each
path, with prediction accuracy peaking at the 300\,ms horizon
(Section~\ref{sec:depth_analysis}).

\subsubsection{Architecture Selection}

The Transformer architecture was selected over \ac{LSTM} approaches
due to parallelisation advantages: \ac{LSTM} networks process
sequences sequentially, requiring $\mathcal{O}(L)$ sequential
operations for sequence length~$L$, whereas Transformer self-attention
computes all token relationships simultaneously in $\mathcal{O}(1)$
sequential depth, reducing inference from approximately 2.5\,s to
approximately 2.5\,ms despite comparable parameter counts. This
selection is validated empirically in
Section~\ref{sec:predictor_selection}.

\subsubsection{Deployed Architecture}

The deployed model uses 3~encoder blocks selected via depth analysis
in Section~\ref{sec:depth_analysis}, each comprising multi-head
self-attention with 5~heads, a position-wise feedforward network of
dimension~360 with GELU activation, layer normalisation, and residual
connections. Dropout rates increase from 10\% to 15\% across depth,
providing implicit regularisation for longer-range predictions
(Table~\ref{tab:transformer_blocks}).

\begin{table}[t]
\centering
\caption{Deployed Transformer block specifications (3-layer, selected
  via depth analysis in Section~\ref{sec:depth_analysis}).}
\label{tab:transformer_blocks}
\begin{tabular}{ccccc}
\toprule
\textbf{Block} & \textbf{FF Dim} & \textbf{Heads} & 
  \textbf{Activation} & \textbf{Dropout} \\
\midrule
1 & 360 & 5 & GELU & 10\% \\
2 & 360 & 5 & GELU & 10\% \\
3 & 360 & 5 & GELU & 15\% \\
\bottomrule
\multicolumn{5}{l}{%
  \footnotesize Dropout increases with depth to regularise 
  longer-range prediction targets.}
\end{tabular}
\end{table}
\subsubsection{Dataset and Preprocessing}

The dataset comprises 30,243,085 samples (3.7\,GB) collected at 1\,ms
intervals from YouTube video streaming sessions over multipath tunnels
under diverse mobility traces. The feature space includes
8~congestion metrics per path: \ac{CWND}, available \ac{CWND}
fraction, bytes in flight, \ac{SRTT}, subflow queue depth, meta queue
depth, lost packets, and delivered packets.

MinMax natural logarithm scaling addresses the heavy-tailed \ac{CWND}
distribution spanning 10--5000 segments, compressing the dynamic
range whilst preserving relative differences critical for rate control
decisions. Data are resampled to uniform 1\,ms intervals and binned
via 500-timestep sliding windows with 100\,ms aggregation. Training
employs an 80--20 train-validation split with early stopping upon
validation loss plateau.

\subsection{DQN-Based Multi-Objective Optimisation}
\label{sec:dqn}

Having obtained forecasts of path characteristics, packets are
scheduled across paths to minimise the impact of mobility-induced link
quality variations on throughput. \ac{DARA} operates as a traffic
splitting algorithm: degradation in Path~1's \ac{QoS} with stable
Path~2 triggers preemptive traffic shifts. The Transformer's
predictions
$Z_p(t) = [\hat{w}_p(t{+}h) \;\; \hat{\tau}_p(t{+}h)]$ are
integrated with current observations
$X_p(t) = [w_p(t) \;\; \tau_p(t)]$ to form the optimisation input
over the set of available subflows~$P$.

\subsubsection{MDP Formulation}

\textbf{Agent:} The multipath scheduler determines per-subflow
\ac{CWND} fractions $\phi_p(t)$ provided to the packet scheduler,
controlling the effective transmission capacity utilised on path~$p$
at time~$t$.

\textbf{Observation and State:} At time~$t$, the agent observes an
18-dimensional state vector~$s_t$:
\begin{equation}
\begin{aligned}
s_t = \Big[&\underbrace{w_p(t),\;\tau_p(t)}_{p \in P},\;
\underbrace{\delta^{\text{mid}}_{w,p},\;
  \delta^{\text{mid}}_{\tau,p}}_{p \in P},\;
\underbrace{\delta^{\text{far}}_{w,p},\;
  \delta^{\text{far}}_{\tau,p}}_{p \in P},\\
&\underbrace{\bar{\phi}_p(t{-}1)}_{p \in P},\;
\underbrace{c^{w}_p,\;c^{\tau}_p}_{p \in P}\Big]
\end{aligned}
\label{eq:state}
\end{equation}
where $w_p(t)$ and $\tau_p(t)$ denote current \ac{CWND} and
\ac{SRTT}; normalised deltas
$\delta^{\text{mid}}_{w,p} = (\hat{w}_p(t{+}h/2) - w_p(t)) /
\max(|w_p(t)|, \epsilon)$ and
$\delta^{\text{far}}_{w,p} = (\hat{w}_p(t{+}h) - w_p(t)) /
\max(|w_p(t)|, \epsilon)$ capture predicted rates of change at
mid-horizon and full-horizon respectively (analogously
for~$\delta_\tau$); $\bar{\phi}_p(t{-}1) = \phi_p(t{-}1)/100$
encodes the previous action; and binary congestion flags
$c^w_p = \mathds{1}[\hat{w}_p(t{+}h) < 0.95 \cdot w_p(t)]$ and
$c^\tau_p = \mathds{1}[\hat{\tau}_p(t{+}h) > 1.1 \cdot \tau_p(t)]$
signal predicted degradation. For $N{=}2$ paths, this yields
$4{+}4{+}4{+}2{+}4 = 18$ dimensions, scaling as~$9N$ for arbitrary
path counts.

\textbf{Action:} The action
$a_t = \{\phi_p(t)\}_{p \in P}$ adjusts \ac{CWND} fractions for all
$N$~paths, where $\phi_p(t) \in \Phi$ draws from a discrete set of
$K$~fraction levels. The joint action space has cardinality~$K^N$.
Section~\ref{sec:action_granularity} presents a sensitivity analysis
comparing 2-level, 3-level, 5-level, and 7-level discretisations,
with the 5-level set
$\Phi = \{30, 47, 65, 82, 100\}\%$ (25~joint actions) selected as
the optimal trade-off between rate control precision and learning
efficiency.

\textbf{Policy:} The policy $\mu(s_t): \mathcal{S} \to \mathcal{A}$
maps states to actions, learnt through \ac{DRL} to maximise
cumulative discounted reward
$\sum_{k=0}^{\infty} \gamma^k r_{t+k}$ with $\gamma = 0.966$.

\subsubsection{Multi-Component Reward Function}

The reward function $R(s_t, a_t)$ comprises six normalised components,
each designed to occupy approximately $\mathcal{O}([-1,1])$ range,
preventing any single objective from dominating gradient updates.
Normalisation is verified empirically in
Section~\ref{sec:reward_analysis}. For each path $p \in P$, let
$w_p(t)$ and $\tau_p(t)$ denote current \ac{CWND} and \ac{SRTT}, and
$\hat{w}_p(t{+}h)$ and $\hat{\tau}_p(t{+}h)$ denote
Transformer-predicted values at horizon~$h$.

\textbf{1.\ Throughput Maximisation ($R_{\text{throughput}}$):}
\begin{equation}
R_{\text{throughput}}(t) = \sum_{p \in P}
  \frac{\hat{w}_p(t{+}h) - w_p(t)}{w_p(t)}
\label{eq:r_throughput}
\end{equation}

\textbf{2.\ Delay Minimisation ($R_{\text{delay}}$):}
\begin{equation}
R_{\text{delay}}(t) = \sum_{p \in P}
  \frac{\tau_p(t) - \hat{\tau}_p(t{+}h)}{\tau_p(t)}
\label{eq:r_delay}
\end{equation}
Negative values signal impending congestion, enabling preemptive rate
reduction.

\textbf{3.\ Preemptive Adaptation ($R_{\text{preemptive}}$):} For
each path $p \in P$:
\begin{equation}
r^{\text{pre}}_p(t) = \begin{cases}
+0.5 & \text{if } \hat{w}_p < w_p \text{ and }
  \phi_p(t) < \phi_p(t{-}1) \\
-0.5 & \text{if } \hat{w}_p < w_p \text{ and }
  \phi_p(t) \geq \phi_p(t{-}1) \\
+0.5 & \text{if } \hat{w}_p \geq w_p \text{ and }
  \phi_p(t) > \phi_p(t{-}1) \\
0 & \text{otherwise}
\end{cases}
\end{equation}
\begin{equation}
R_{\text{preemptive}}(t) = \frac{1}{N}
  \sum_{p \in P} r^{\text{pre}}_p(t)
\label{eq:r_preemptive}
\end{equation}

\textbf{4.\ Stability Penalty ($R_{\text{stability}}$):}
\begin{equation}
R_{\text{stability}}(t) = -\frac{1}{100}
  \sum_{p \in P} |\phi_p(t) - \phi_p(t{-}1)|
\label{eq:r_stability}
\end{equation}

\textbf{5.\ Quality Sustainability ($R_{\text{quality}}$):}
\begin{equation}
R_{\text{quality}}(t) = \frac{1}{10^6} \sum_{p \in P}
  \frac{\hat{w}_p(t{+}h)}{\max(1, \hat{\tau}_p(t{+}h))}
  \times 8 \times 1500
\label{eq:r_quality}
\end{equation}
Exponentially smoothed
($\bar{R}_{\text{quality}}(t) = 0.9 \cdot
\bar{R}_{\text{quality}}(t{-}1) + 0.1 \cdot R_{\text{quality}}(t)$)
to prevent oscillatory policies.

\textbf{6.\ Idleness Prevention ($R_{\text{idleness}}$):}
\begin{equation}
R_{\text{idleness}}(t) = \begin{cases}
-1 & \text{if } \phi_p(t) = \phi_{\min} \;\forall\, p \in P \\
0 & \text{otherwise}
\end{cases}
\label{eq:r_idleness}
\end{equation}
where $\phi_{\min} = 30\%$.

The six components couple to $\phi$ adjustments through two pathways.
\textit{Direct coupling}: the stability component penalises
$|\phi_p(t) - \phi_p(t{-}1)|$ instantaneously; idleness fires when
all $\phi_p = \phi_{\min}$; and the preemptive component provides
explicit reward shaping by comparing $\phi_p(t)$ with
$\phi_p(t{-}1)$ conditional on the predicted \ac{CWND} direction.
\textit{Indirect coupling}: throughput, delay, and quality depend on
future \ac{CWND} and \ac{SRTT} values influenced by the current
$\phi$ choice through its effect on in-flight segment counts and
congestion dynamics. The discount factor $\gamma = 0.966$ assigns
35\% weight to rewards approximately 30~steps (3\,s) ahead.


\subsubsection{Reward Aggregation}
\label{sec:reward_aggregation}

The combined reward is:
\begin{equation}
R(s_t, a_t) = \sum_{x \in \mathcal{W}}
  w_x \cdot R_x
\label{eq:reward}
\end{equation}
where $\mathcal{W} = \{\text{tput}, \text{delay}, \text{qual},
\text{pre}, \text{stab}, \text{idle}\}$. All components are
normalised to $\mathcal{O}([-1,1])$ before weighting, ensuring
weights encode relative objective importance rather than compensating
for scale mismatches; empirically verified ranges are reported in
Section~\ref{sec:reward_analysis}. Weight calibration is described
in Section~\ref{sec:reward_weights}; the discovered configuration
(Table~\ref{tab:final_config}) assigns highest weight to throughput
and second-highest to the preemptive component, confirming that
explicit anticipatory reward shaping is necessary when components
are properly normalised. Stability functions as a soft regulariser
with the lowest assigned weight.

\subsubsection{Conflict Resolution}
\label{sec:conflict_resolution}

Three principal conflict scenarios arise under weighted linear
scalarisation (Equation~\ref{eq:reward}):

\textbf{Throughput versus stability.} Exploiting a predicted burst
requires rapid $\phi$~increase, incurring stability penalty. The
weight ratio
$w_{\text{throughput}}/w_{\text{stability}} \approx 19$ ensures
throughput gains dominate except for very large $\phi$~jumps:
\begin{equation}
\Delta\phi_p > 0 \iff w_{\text{tput}} \cdot
  \frac{\hat{w}_p - w_p}{w_p} > w_{\text{stab}} \cdot
  \frac{|\Delta\phi_p|}{100}
\end{equation}

\subsubsection{Reward Weight Calibration}
\label{sec:reward_weights}

Manual tuning of the six weights is prohibitive due to non-convex
interactions. A random search evaluates candidate weight vectors
sampled from the ranges: $w_{\text{throughput}},
w_{\text{delay}} \in [0.1, 5.0]$; $w_{\text{quality}},
w_{\text{preemptive}} \in [0.01, 3.0]$; $w_{\text{stability}}
\in [0.01, 2.0]$; $w_{\text{idleness}} \in [0.01, 1.0]$. Each
candidate is scored via $S = \bar{R} + 10 \cdot P_{\text{preemptive}}$,
balancing mean episode reward with the fraction of congestion events
handled preemptively. The discovered weights are reported in
Table~\ref{tab:final_config}.

\textbf{Preemptive versus idleness.} When both paths show predicted
degradation, the preemptive reward encourages $\phi$~reduction on
both, potentially triggering idleness penalty. In practice, the
throughput loss from dual-minimum allocation outweighs the preemptive
gain, confirmed by the low frequency of minimum-minimum actions
observed in training (Section~\ref{sec:ablation}).

\textbf{Delay versus throughput.} The weight ratio
$w_{\text{throughput}}/w_{\text{delay}} \approx 2.5$ resolves this
in favour of throughput, accepting transient delay during burst
phases. The \ac{DQN} learns to increase $\phi_p$ when:
\begin{equation}
w_{\text{tput}} \cdot \frac{\hat{w}_p - w_p}{w_p} >
  w_{\text{delay}} \cdot \frac{\hat{\tau}_p - \tau_p}{\tau_p}
\label{eq:conflict_threshold}
\end{equation}

These resolutions operate through gradient-based policy optimisation:
the \ac{DQN} learns action-value estimates internalising the weighted
trade-offs across diverse training conditions.

\subsection{Hyperparameter Optimisation}
\label{sec:nas}


Optuna-based \ac{NAS} with Tree-structured Parzen Estimator sampling
and median pruning explored the DQN architectural hyperparameter space
over 100~trials on a high-performance computing cluster (3,500 cores,
30\,TB RAM), totalling approximately 350~compute-hours. The search
spans temporal resolution and DQN architecture: hidden dimension,
layer count, learning rate, batch size, $\gamma$, $\tau$, replay
capacity, and exploration decay. Reward weights are calibrated
separately via random search over 50~iterations as described in
Section~\ref{sec:reward_weights}. Each trial trains a \ac{DQN} agent
on 500,000-row samples from the 30M-sample dataset, evaluated via a
composite score balancing average reward (40\%), inverse delay (20\%),
and convergence speed (10\%). Table~\ref{tab:final_config} summarises
the full deployed configuration.

\begin{table}[t]
\centering
\caption{Final DARA deployed configuration.}
\label{tab:final_config}
\begin{tabular}{ll}
\toprule
\textbf{Component} & \textbf{Value} \\
\midrule
\multicolumn{2}{l}{\textit{Prediction model}} \\
Transformer depth      & 3 layers \\
Attention heads        & 5 \\
Feedforward dimension  & 360 \\
Input window           & 8 $\times$ 100\,ms steps \\
Prediction horizons    & 5 $\times$ 100\,ms (100--500\,ms) \\
\midrule
\multicolumn{2}{l}{\textit{DQN policy}} \\
State dimensions       & 18 (Equation~\ref{eq:state}) \\
Action set $\Phi$      & $\{30,47,65,82,100\}\%$ (25 joint) \\
Hidden units / layers  & 256 / 2 \\
Learning rate $\alpha$ & $1.25 \times 10^{-4}$ \\
Discount $\gamma$      & 0.966 \\
Soft update $\tau$     & 0.006 \\
Batch size / Memory    & 128 / 15,000 \\
$\epsilon$ start / end / decay & 0.7 / 0.1 / 900 steps \\
\midrule
\multicolumn{2}{l}{\textit{Reward weights}} \\
$w_{\text{throughput}}$  & 3.329 \\
$w_{\text{preemptive}}$  & 2.194 \\
$w_{\text{quality}}$     & 2.130 \\
$w_{\text{delay}}$       & 1.332 \\
$w_{\text{idleness}}$    & 0.974 \\
$w_{\text{stability}}$   & 0.175 \\
\bottomrule
\end{tabular}
\end{table}

\subsection{Training and Execution Algorithm}
\label{sec:algorithm}

Algorithm~\ref{alg:dqn_dara} presents \ac{DARA}'s training loop.
States~$s_t$ comprise the 18-dimensional vector defined in
Equation~\ref{eq:state}, formed by concatenating normalised current
metrics, predicted deltas at mid-horizon and full-horizon, previous
actions, and congestion flags derived from frozen Transformer
predictions. Experience tuples $(s_t, a_t, r_t, s_{t+1})$ populate
replay buffer~$D$ with capacity~$C$. Once $|D| \geq B$, mini-batch
updates compute temporal-difference targets using the target network,
with the policy network updated via gradient descent on
$\mathcal{L} = (1/|\mathcal{B}|) \sum_j
(Q_\theta(s_j, a_j) - y_j)^2$. The $\epsilon$-greedy exploration
with exponential decay ($\epsilon_{\text{start}} = 0.7$,
$\epsilon_{\text{end}} = 0.1$, 900-step decay) reflects that
Transformer predictions provide informative state representations
from training onset. After training, the frozen policy network and
Transformer are deployed to the userspace controller, where the
100\,ms control cycle produces \ac{CWND} fraction updates applied
by the kernelspace scheduler at ${\sim}$1\,ms packet dispatch
granularity.

\begin{algorithm}[t]
\caption{DARA Training with Transformer-Informed State}
\label{alg:dqn_dara}
\begin{algorithmic}[1]
\State \textbf{Load} pre-trained Transformer $\mathcal{M}$
\State \textbf{Initialise} policy network $Q_\theta$, target network
  $Q_{\theta'}$ with $\theta' \leftarrow \theta$
\State \textbf{Initialise} replay buffer $D$ with capacity $C$
\For{episode $e = 1$ to $N_{\text{episodes}}$}
  \State Collect current telemetry $H_t$ for all paths $p \in P$
  \State Obtain predictions:
    $\hat{Z}_p \leftarrow \mathcal{M}(H_t)$ for all $p \in P$
  \State Construct state $s_t$ via Equation~\ref{eq:state}
  \While{state $s_t$ is not terminal or truncated}
    \State Compute $\epsilon$ from decay schedule
    \If{$\text{rand}() < \epsilon$}
      \State Select random action $a_t \in \Phi^N$
    \Else
      \State $a_t = \arg\max_{a'} Q_\theta(s_t, a')$
    \EndIf
    \State Execute $a_t = \{\phi_p(t)\}_{p \in P}$ to the scheduler
    \State Advance to $t{+}1$, collect $H_{t+1}$, predict
      $\hat{Z}_p(t{+}1)$
    \State Compute $r_t \leftarrow R(s_t, a_t)$ via
      Equation~\ref{eq:reward}
    \State Construct $s_{t+1}$ via Equation~\ref{eq:state}
    \State Store $(s_t, a_t, r_t, s_{t+1})$ in $D$
    \If{$|D| \geq B$}
      \State Sample mini-batch $\mathcal{B}$ from $D$
      \State $y_j = r_j + \gamma \max_{a''} Q_{\theta'}(s'_j, a'')$
      \State Update $\theta$ by minimising
        $\frac{1}{|\mathcal{B}|}\sum_j
        \mathcal{L}(Q_\theta(s_j, a_j), y_j)$
      \State $\theta' \leftarrow \tau\theta + (1{-}\tau)\theta'$
    \EndIf
    \State $s_t \leftarrow s_{t+1}$
  \EndWhile
\EndFor
\end{algorithmic}
\end{algorithm}

\section{Testing Methodology}
\label{sec:methodology}

This section evaluates \ac{DARA}'s performance within the
\ac{MP-DCCP} multipath framework using a Mininet testbed with traces
from real moving users. \ac{DARA} is compared against fixed
rule-based and learning-based schedulers under realistic volatile
network conditions.

\subsection{Mininet Architecture}

Mininet\footnote{http://mininet.org/} provides network emulation
through lightweight virtualisation using Linux network namespaces,
executing real kernel code, switch implementations, and unmodified
applications. Unlike discrete-event simulators, Mininet runs genuine
protocol implementations, including Deutsche Telekom's out-of-tree
\ac{MP-DCCP} kernel module, ensuring observed performance reflects
actual transport dynamics.

\subsection{MP-DCCP and Protocol-Agnostic Operation}

Although \ac{MP-TCP} was the dominant multipath protocol for some
time, its reliable transmission makes it less favoured for future
networks due to the increase in video-on-demand consumption (estimated
to be 80\% of mobile traffic by 2029~\cite{ericsson_2022}).
Unreliable multipath frameworks instead encapsulate original IP
packets within the multipath transport protocol, enabling
protocol-agnostic operation. \ac{MP-QUIC} and \ac{MP-DCCP} both
support these features as well as \ac{ATSSS} capabilities. \ac{MP-DCCP}
has a more developed
implementation\footnote{https://github.com/telekom/mp-dccp/tree/mpdccp\_v03\_k4.14}
and is therefore preferred. This encapsulation ensures \ac{DARA}
requires no protocol-specific adjustments: the scheduler controls only
the outer transport layer whilst encapsulated TCP, QUIC, or UDP
traffic operates unmodified.

\subsection{Replaying Volatile Path Traffic}

The Mahi-Mahi\footnote{http://mahimahi.mit.edu/} framework provides
deterministic trace replay of cellular capacity variations.
Mahi-Mahi consumes packet-delivery traces, timestamped records of
MTU-sized transmission opportunities captured from saturated links of
moving users, and enforces identical packet delivery sequences across
experimental runs. Given trace
$T = \{(t_1, s_1), (t_2, s_2), \ldots\}$ where $t_i$ denotes the
timestamp and $s_i$ the deliverable bytes, transmission of segment~$k$
proceeds only when
$\sum_{j \leq k} |segment_j| \leq \sum_{t_i \leq t_{\text{now}}} s_i$.

\begin{figure}[t]
\centering
\includegraphics[width=\linewidth]{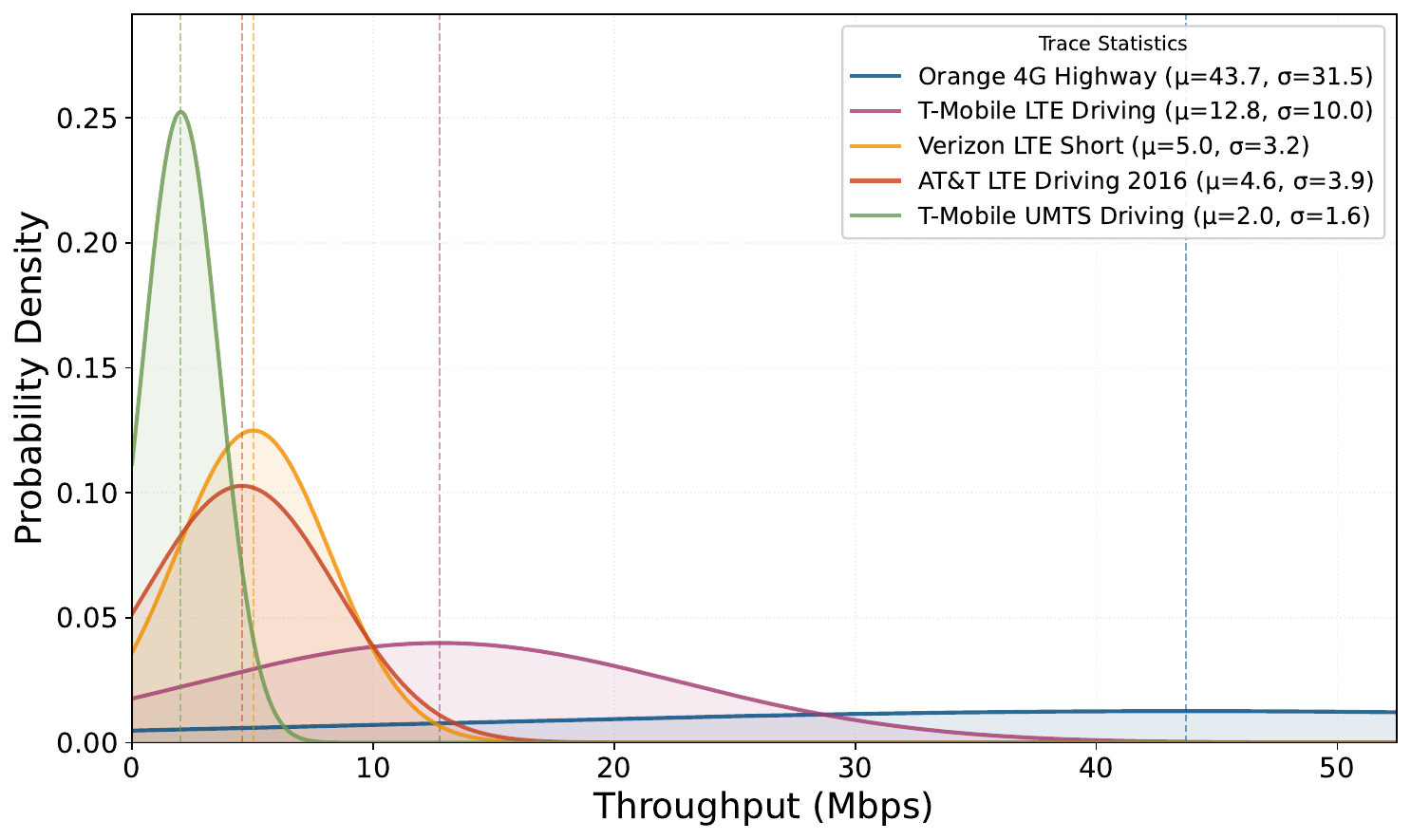}
\caption{Path throughput normal distribution.}
\label{fig:Distribution}
\end{figure}

Figure~\ref{fig:Distribution} presents the throughput distributions of
five publicly available cellular
traces\footnote{https://github.com/joelromanky/mahimahi/tree/master/traces}
used in evaluation. Captured from moving vehicles experiencing
continuous handovers and signal fluctuations, these traces exhibit
standard deviations exceeding 75\% of mean throughput.

\subsection{Definition of Comparing Algorithms}

The performance of \ac{DARA} is compared with scheduling algorithms
selected to represent the spectrum of multipath scheduling paradigms:
static rule-based approaches (\ac{RR}, \ac{CPF}), path-estimation
heuristics (\ac{OTIAS}, \ac{ACPF}), and learning-based predictive
methods (\ac{BPF}, Peekaboo, \ac{BLEST}). Each scheduler operates
within the same Mininet testbed under identical network conditions.

\subsubsection{Cheapest Pipe First (CPF)}
\ac{CPF} prioritises data transmission on the path with the highest
priority, using secondary paths once the primary path exhausts its
\ac{CWND}~\cite{amend2019framework}. When path priorities are equal,
\ac{CPF} transitions to \ac{SRTT} mode.

\subsubsection{Best Path First (BPF)}
\ac{BPF}~\cite{maglione2025intelligent} employs \ac{LSTM} networks to
predict future path quality over a 5-second horizon, achieving 20\%
delay reduction and 30\% faster FTP downloads in high-mobility
scenarios. The fixed 5-second horizon may be suboptimal for
sub-second burst dynamics.

\subsubsection{Adaptive Cheapest Pipe First (ACPF)}
\ac{ACPF}~\cite{10.1145/3472305.3472316} extends \ac{CPF} with
dynamic congestion window adjustment, though it remains fundamentally
reactive.

\subsubsection{Out-of-Order Transmission for In-order Arrival (OTIAS)}
\ac{OTIAS}~\cite{6844729} deliberately transmits packets out-of-order
to achieve in-order arrival at the receiver, but assumes stable
\ac{RTT} measurements, causing mispredictions during handovers.

\subsubsection{Peekaboo}
Peekaboo~\cite{9110610} leverages multi-armed bandit theory for
dynamic heterogeneous paths but is unable to exploit long-range
temporal patterns.

\subsubsection{BLocking ESTimate (BLEST)}
\ac{BLEST}~\cite{7497206} addresses \ac{HoL} blocking through
predictive blocking estimation, with heuristic burst detection that
degrades under rapid oscillation.

\subsection{Test Scenarios}

\subsubsection{File Transfer (FTP/QUIC)}
Completion time for 150\,MB file transfers is measured, recording
per-packet delay distributions and effective throughput averaged over
100\,ms windows.

\subsubsection{YouTube Streaming}
YouTube playback with 200-second seeks every 20~seconds prevents
buffer accumulation masking scheduler behaviour. Metrics include
\ac{APR}, rebuffering frequency, quality switch rate, and initial
delay.

\subsubsection{Live Video Streaming}
Live streaming via HLS operates with approximately 0.5\,s buffering,
transforming capacity mismatches into immediate rebuffering events.

\section{Experimental Results and Discussion}
\label{sec:results}

This section presents comprehensive evaluation of \ac{DARA}'s
predictive rate allocation mechanism. The analysis proceeds in two
stages: first, design validation (Section~\ref{sec:design_validation})
isolates the contribution of each architectural decision through
ablation studies conducted on the training dataset; second, controlled
burst scenario validation (Section~\ref{sec:controlled_burst}) and
real-world trace evaluation (Section~\ref{sec:realworld_results})
assess the integrated system under deterministic burst patterns and
realistic high-mobility traces respectively.

\subsection{Design Validation and Ablation Studies}
\label{sec:design_validation}

This section presents systematic evaluation of DARA's architectural
decisions through five analyses: predictor architecture selection,
Transformer depth optimisation, reward component normalisation
verification, action-space granularity sensitivity, and full component
ablation with statistical validation. All experiments use the dataset
described in Section~\ref{sec:transformer}, with multiple~independent runs
per configuration unless otherwise stated.

\subsection{Predictor Architecture Selection}
\label{sec:predictor_selection}

Four candidate architectures are evaluated: Transformer (3-block
attention), LSTM (2-layer, 128-unit recurrent), MLP (2-layer,
256-unit feedforward with flattened input), and a linear baseline
(single-layer projection). All models receive identical 8-step input
sequences of 90~features and produce 20-dimensional outputs (predicted
\ac{CWND} and \ac{SRTT} at 5~horizons for 2~paths).

Table~\ref{tab:predictor_comparison} presents the evaluation. The
Transformer achieves the lowest \ac{NRMSE}, followed by the linear
model, LSTM, and MLP. The linear model's competitive accuracy reflects
substantial autocorrelation in short-horizon congestion metrics; the
MLP performs worst as its flattened input destroys temporal ordering.
For downstream \ac{DQN} performance, the Transformer, linear, and
LSTM predictors yield comparable normalised rewards, indicating
moderate prediction errors do not catastrophically degrade policy
learning. All architectures operate within the 100\,ms budget, with
the Transformer offering the best accuracy-to-latency ratio at
approximately 305K~parameters and 1.8\,ms inference.

\begin{table}[t]
\centering
\caption{Predictor architecture comparison. NRMSE: normalised root
  mean square error; Reward: normalised downstream DQN performance
  (Transformer = 1.00).}
\label{tab:predictor_comparison}
\begin{tabular}{lcccc}
\toprule
\textbf{Architecture} & \textbf{NRMSE} & \textbf{Params} &
  \textbf{Latency (ms)} & \textbf{Reward} \\
\midrule
Transformer (3L) & 0.0009 & 305K & 1.8 & 1.00 \\
Linear            & 0.0022 & 2K   & 0.0 & 0.97 \\
LSTM (2L)         & 0.0031 & 247K & 0.2 & 0.97 \\
MLP (2L)          & 0.0376 & 291K & 0.1 & ---$^{*}$ \\
\bottomrule
\multicolumn{5}{l}{\footnotesize $^{*}$Excluded: unclamped
  inverse-transform artefact.}
\end{tabular}
\end{table}

\subsection{Transformer Depth Analysis}
\label{sec:depth_analysis}

Given the Transformer's selection, encoder depths of 1, 2, 3, 4, and
6~layers are evaluated. A non-monotonic relationship between depth and
accuracy emerges: error decreases from 1 to 3~layers, then increases
for 4 and~6, as deeper models overfit to training-set temporal
patterns that do not generalise to unseen mobility traces. At the
300\,ms horizon most relevant to preemptive scheduling, the 3-layer
model achieves the lowest error across all depths. The 3-layer
configuration is selected as it achieves the highest combined
accuracy-efficiency score whilst remaining within the real-time
inference budget ($\approx$2.5\,ms, less than 3\% of the 100\,ms
control cycle). The speedup over \ac{LSTM} derives from the
architectural shift to parallel self-attention; all evaluated depths
satisfy the inference budget.

\subsection{Reward Component Normalisation}
\label{sec:reward_analysis}

A critical design requirement is that reward components occupy
comparable numerical ranges, ensuring weight values reflect genuine
objective trade-offs. The quality component divides by~$10^6$ to
convert raw bitrate estimates to unit-scale values, and the stability
component divides raw $\phi$~differences by~100. Empirical
verification over representative training episodes confirms all
components fall within $\mathcal{O}([-1, 1])$: throughput
$[-0.014, 0.016]$, delay $[-0.007, 0.009]$, preemptive $[-1.0,
1.0]$, stability $[-1.22, 0.0]$, quality $[0.013, 0.024]$, and
idleness $[-1.0, 0.0]$. The preemptive component exhibits the widest
dynamic range, providing strong training gradients; throughput and
delay have small absolute magnitudes but high temporal variation,
encoding rapid congestion dynamics.

\subsection{Action Space Granularity}
\label{sec:action_granularity}

Four configurations are evaluated: 2-level $\{30, 100\}\%$
(4~joint actions), 3-level $\{30, 65, 100\}\%$ (9), 5-level
$\{30, 47, 65, 82, 100\}\%$ (25), and 7-level
$\{30, 42, 53, 65, 77, 88, 100\}\%$ (49). Each is trained for
300~episodes across 3~runs.

Table~\ref{tab:action_granularity} presents the results. Reward
increases substantially from 2-level to 5-level, then plateaus at
7-level. The 2-level configuration constrains the policy to binary
full/minimum allocation, preventing graduated responses to moderate
capacity changes. The 7-level configuration achieves marginally lower
mean reward than 5-level with substantially higher variance
($\pm 16.8$ vs $\pm 3.8$), reflecting insufficient exploration of the
49-action space within the training budget. The 5-level configuration
is selected as the optimal trade-off between rate control precision
and learning efficiency.

\begin{table}[t]
\centering
\caption{Action granularity analysis. Reward: mean $\pm$ std over
  3~runs; Preemptive: fraction of congestion events handled
  anticipatorily.}
\label{tab:action_granularity}
\begin{tabular}{lccc}
\toprule
\textbf{Configuration} & \textbf{Actions} & \textbf{Reward} &
  \textbf{Preemptive (\%)} \\
\midrule
2-Level & 4  & $17.1 \pm 3.9$   & 53.4 \\
3-Level & 9  & $101.5 \pm 5.4$  & 56.9 \\
5-Level & 25 & $155.0 \pm 3.8$  & 63.6 \\
7-Level & 49 & $152.6 \pm 16.8$ & 62.4 \\
\bottomrule
\end{tabular}
\end{table}

\subsection{Component Ablation and Statistical Validation}
\label{sec:ablation}

The full DARA system is compared against six baselines: random
$\phi$~selection, reactive adjustment (increase/decrease by 10\%
based on observed \ac{CWND} trends), three static allocations
($\phi = 30\%$, $65\%$, $100\%$), DQN without Transformer predictions
(zero-vector state), and Ground-Truth DQN (ground-truth future values).
Each configuration is evaluated over multiple~independent runs.

\begin{figure}[t]
\centering
\includegraphics[width=\linewidth]{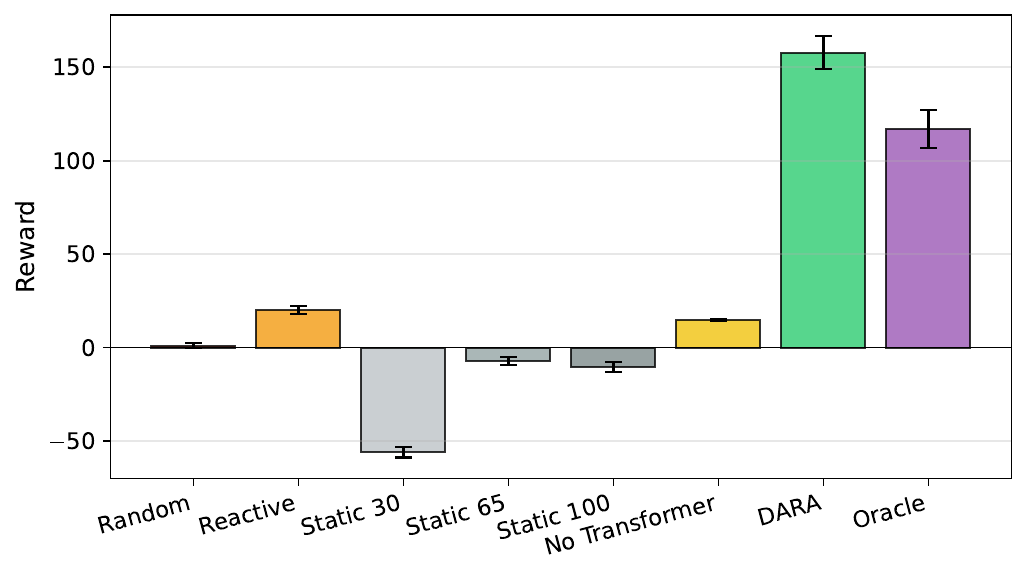}
\caption{Mean reward with 95\% confidence intervals across all
  methods.}
\label{fig:ablation_results}
\end{figure}

Figure~\ref{fig:ablation_results} presents mean rewards with 95\% confidence intervals. DARA achieves the highest reward (157.9). The no-Transformer ablation confirms that prediction is essential: without forecasts, the DQN defaults to near-constant allocation with zero preemptive rate, degenerating to a static scheduler. The Ground-Truth DQN, supplied with raw kernel telemetry in place of Transformer predictions, achieves a higher preemptive rate (73.1\%) but lower overall reward (116.9), as high-frequency measurement noise in the raw state degrades policy consistency relative to the Transformer's smoothed representations. DARA handles 62.5\% of congestion events through anticipatory $\phi$~reduction.

Table~\ref{tab:ablation_results} summarises the performance of all
methods. Table~\ref{tab:statistical_tests} presents the statistical
validation. All pairwise comparisons achieve significance under both
Welch's $t$-test and Mann-Whitney~$U$ test ($p < 10^{-5}$), with
large effect sizes (Cohen's $d$) for all comparisons. Bootstrap 95\%
confidence intervals for reward differences all exclude zero,
confirming robust improvements.

\begin{table}[t]
\centering
\caption{Ablation results. $R$: mean reward; Pre.: preemptive rate.}
\label{tab:ablation_results}
\begin{tabular}{lcc}
\toprule
\textbf{Method} & $R$ & \textbf{Pre.\ (\%)} \\
\midrule
Random          &   $1.0$  &  0.0 \\
Reactive        &  $20.0$  &  0.0 \\
Static 30       & $-56.0$  &  0.0 \\
Static 65       &  $-7.2$  &  0.0 \\
Static 100      & $-10.4$  &  0.0 \\
No Transformer  &  $14.7$  &  0.0 \\
Ground-Truth DQN          & $116.9$  & 73.1 \\
\textbf{DARA}   & $\mathbf{157.9}$ & $\mathbf{62.5}$ \\
\bottomrule
\end{tabular}
\end{table}

\begin{table}[t]
\centering
\caption{Statistical comparison of DARA against baselines.
  $\Delta R$: mean reward difference; $d$: Cohen's $d$;
  $p$: Welch's $t$-test; CI: bootstrap 95\% confidence interval.}
\label{tab:statistical_tests}
\begin{tabular}{lcccc}
\toprule
\textbf{Comparison} & $\Delta R$ & $d$ & $p$ & \textbf{95\% CI} \\
\midrule
vs.\ Random             & 156.9 & 23.5 & $<10^{-7}$ & [151.7, 163.4] \\
vs.\ Reactive           & 137.9 & 12.2 & $<10^{-7}$ & [132.6, 144.0] \\
vs.\ Static 65          & 165.1 & 16.3 & $<10^{-7}$ & [159.6, 171.7] \\
vs.\ Static 100         & 168.2 & 13.7 & $<10^{-8}$ & [162.9, 174.7] \\
vs.\ No Transformer     & 143.1 & 28.4 & $<10^{-6}$ & [138.4, 149.5] \\
vs.\ Ground-Truth DQN   &  40.9 &  5.4 & $<10^{-5}$ & [32.5, 49.5] \\
\bottomrule
\end{tabular}
\end{table}

\subsection{Controlled Burst Scenario Validation}
\label{sec:controlled_burst}

\begin{figure}[t]
\centering
\includegraphics[width=\linewidth]{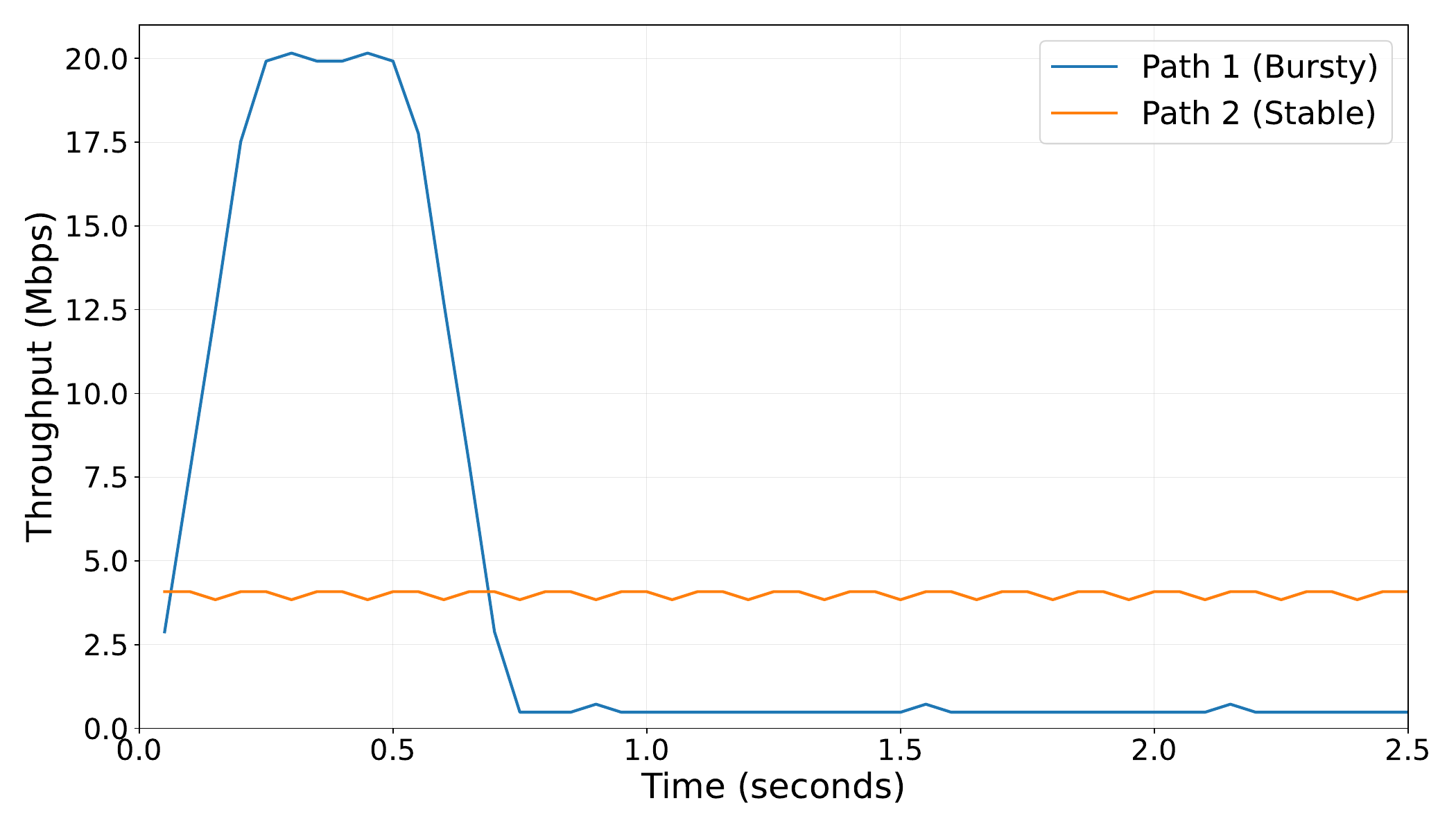}
\caption{Asymmetric burst pattern showing one complete 2.5-second
  cycle. Path~1 exhibits 700\,ms bursts peaking at 10\,Mbps,
  collapsing to 0.2\,Mbps baseline; Path~2 maintains stable
  1.8\,Mbps throughout.}
\label{fig:asymmetric_burst}
\end{figure}

To isolate \ac{DARA}'s predictive burst exploitation capability, a
synthetic asymmetric burst pattern was designed to create reproducible
conditions challenging scheduler responsiveness. As illustrated in
Figure~\ref{fig:asymmetric_burst}, Path~1 exhibits 700\,ms bursts
peaking at 10\,Mbps with 2.5-second cycles, collapsing to 0.2\,Mbps
between bursts; Path~2 maintains stable 1.8\,Mbps throughout. This
produces combined availability ranging from 11.8\,Mbps during burst
peaks to 2.0\,Mbps during troughs, a $5.9\times$ capacity fluctuation
exposing fundamental differences between predictive and reactive
scheduling architectures. The 700\,ms burst duration is sufficiently
long for the Transformer to observe initial capacity changes and
predict burst continuation within its horizon, and the $50\times$
capacity ratio between burst peak and baseline creates conditions
where allocation errors manifest as measurable throughput degradation.

\subsubsection{File Transfer Performance}

\begin{figure}[t]
\centering
\includegraphics[width=\linewidth]{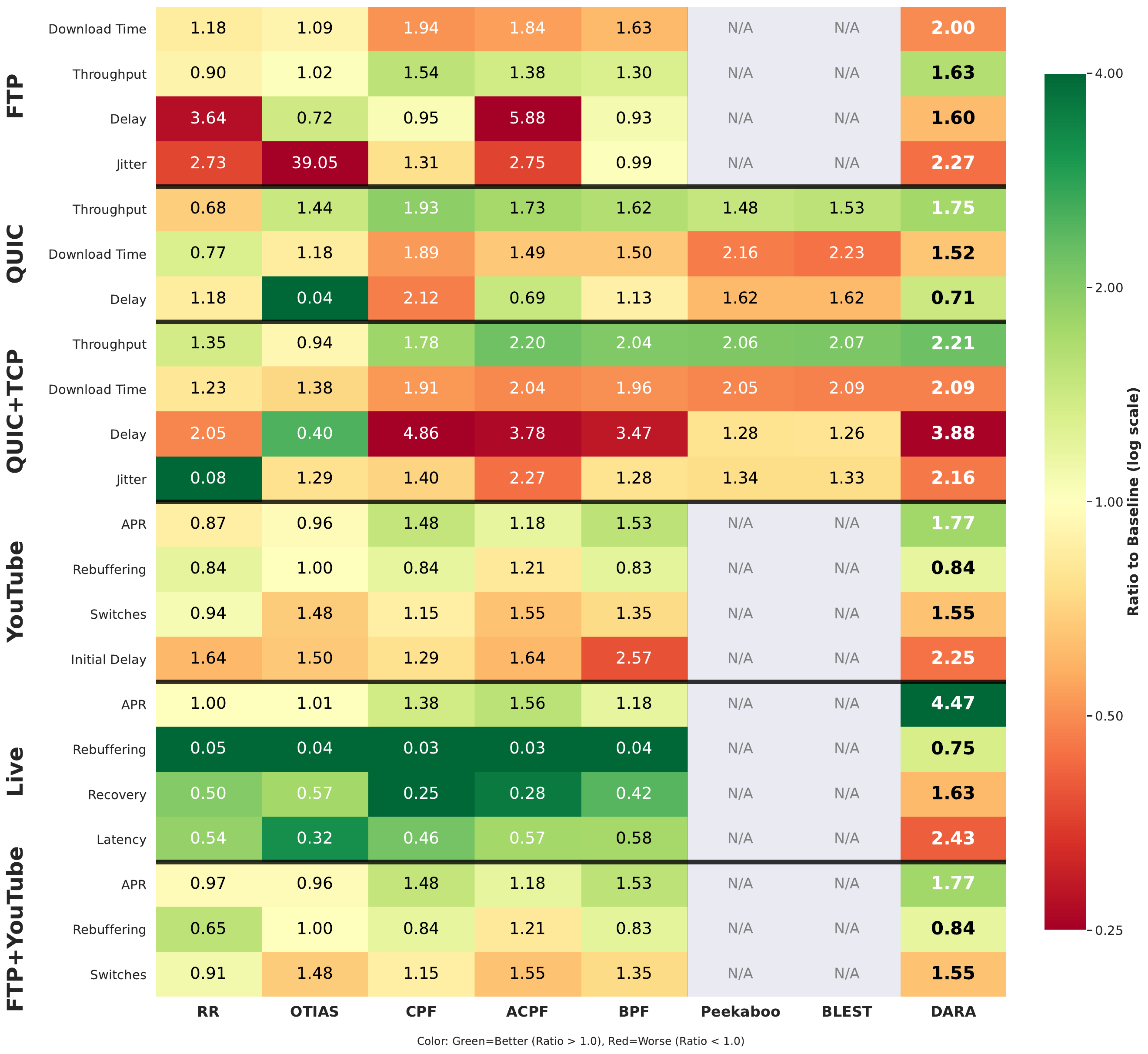}
\caption{Normalised performance metrics for controlled burst scenario.
  Values $>1.0$ indicate improvement over single-path baseline.
  \ac{DARA} achieves consistent leadership across applications, with
  advantages amplifying as latency sensitivity increases
  (FTP $\to$ YouTube $\to$ Live).}
\label{fig:Special_Case}
\end{figure}

Figure~\ref{fig:Special_Case} presents normalised performance metrics
across all evaluated schedulers and application types, with values
exceeding 1.0 indicating improvement over the single-path baseline.
\ac{DARA} achieves consistent throughput leadership across all
transfer types, with gains amplifying for mixed QUIC+TCP traffic.

\begin{figure}[t]
\centering
\includegraphics[width=\linewidth]{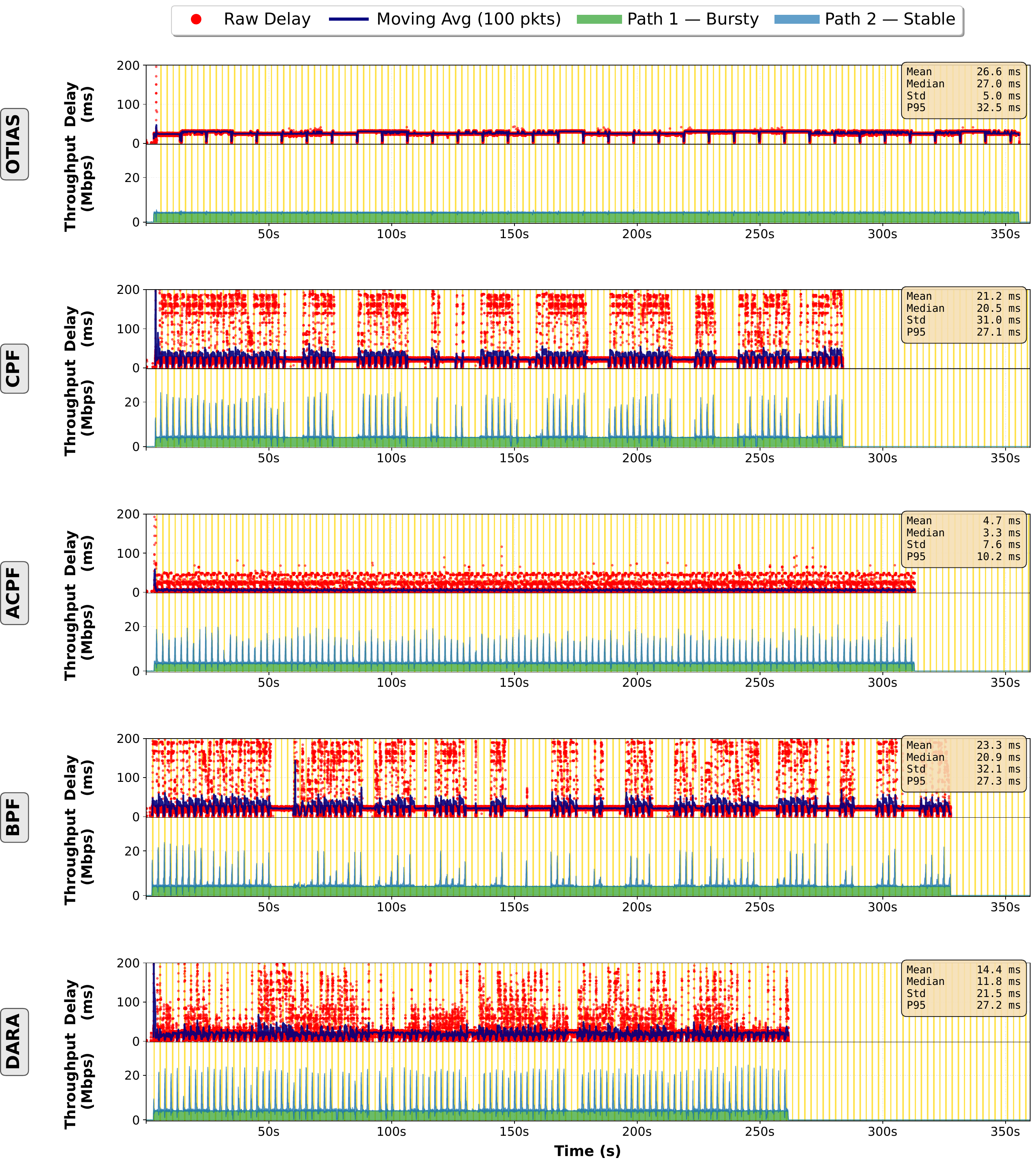}
\caption{Per-scheduler time-series showing delay (top, per-packet raw
  and 100-packet moving average) and throughput (bottom, per-path)
  for FTP under the asymmetric burst pattern. \ac{DARA} (bottom
  panel) maintains stable low delay with sustained burst
  exploitation, whilst reactive schedulers exhibit periodic delay
  spikes aligned with burst-trough transitions.}
\label{fig:asymmetric_time}
\end{figure}

The throughput differential between \ac{DARA} and reactive schedulers
stems from superior burst window utilisation, visible in
Figure~\ref{fig:asymmetric_time}. During burst phases, \ac{DARA}'s
Transformer-informed \ac{DQN} maintains elevated \ac{CWND} fraction
allocation to Path~1. Reactive schedulers exhibit
observation-reaction lag, continuing primary path allocation based on
historical \ac{CWND} measurements until congestion feedback arrives
tens of milliseconds after capacity collapse.

\ac{DARA} achieves the lowest mean and median delay among multipath
schedulers, with a tight distribution and a stable moving-average
baseline throughout, confirming that predictive scheduling eliminates
the periodic delay spikes caused by burst-trough transition
mismanagement. \ac{CPF} exhibits mean and median delays substantially
higher than \ac{DARA}, with periodic clustering aligned with the
burst cycle. \ac{ACPF} achieves the lowest absolute delay through
aggressive queue-based throttling but at a notable throughput cost,
illustrating the trade-off that the predictive architecture resolves.
Among learning-based schedulers, Peekaboo and \ac{BLEST} both fall
short of \ac{DARA}'s throughput gain, whilst path-agnostic \ac{RR}
degrades below single-path baseline.

\subsubsection{Streaming Application Performance}

\ac{DARA} achieves the highest YouTube \ac{APR} improvement,
exceeding \ac{BPF}, \ac{CPF}, and \ac{ACPF}. Live streaming with
0.5-second buffers exposes fundamental limitations of non-predictive
approaches: \ac{DARA} achieves substantial \ac{APR} improvement and a
25\% rebuffering reduction, whilst all other schedulers exhibit
near-continuous stalling.

\subsection{Real-World Trace Evaluation}
\label{sec:realworld_results}

Five cellular traces captured from moving vehicles test whether
\ac{DARA}'s advantages persist under stochastic timing and irregular
capacity patterns. All metrics are reported as ratios versus \ac{CPF}.

\subsubsection{File Transfer Performance}

\begin{figure}[!htp]
\centering
\includegraphics[width=0.95\linewidth]{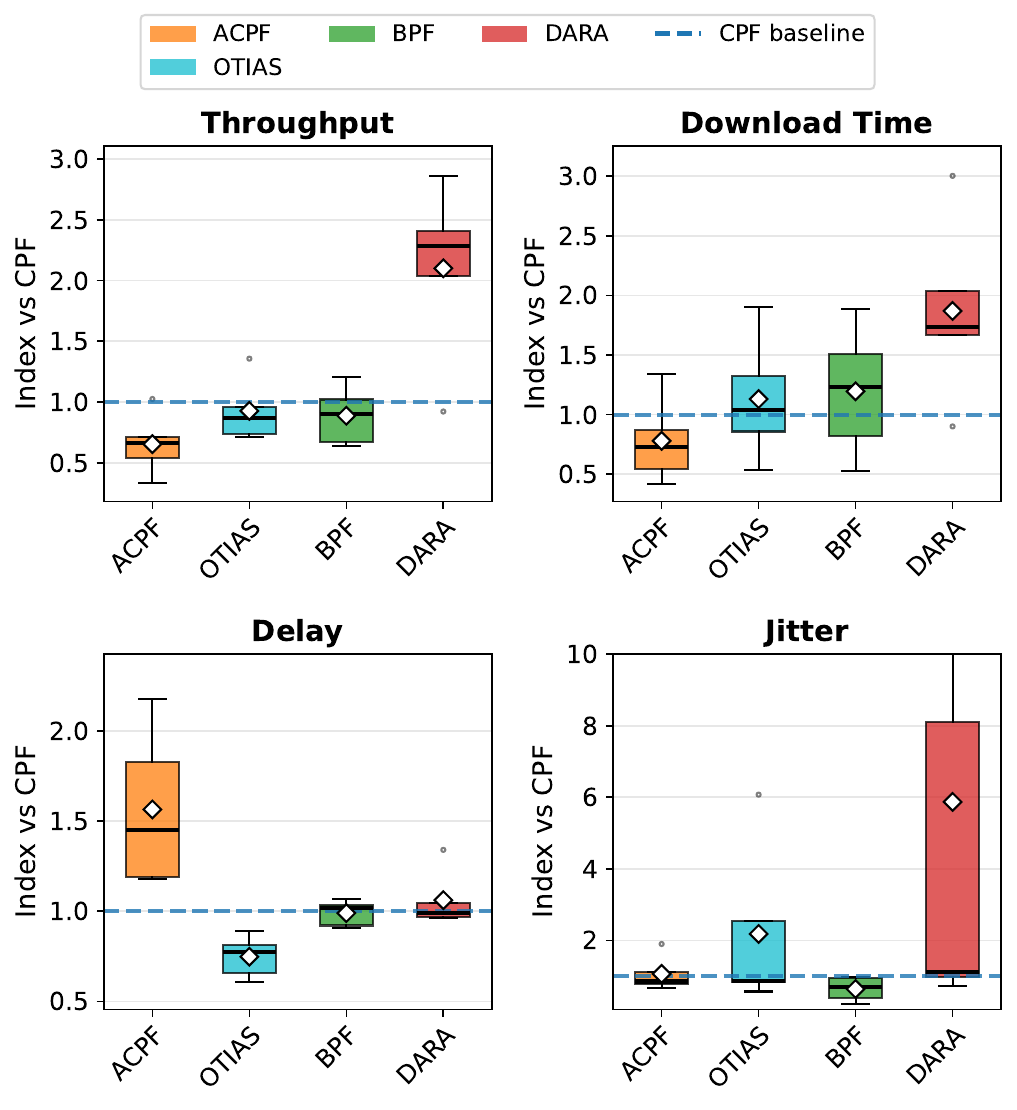}
\caption{TCP file transfer performance distributions across five cellular traces (FTP application protocol; diamond markers indicate the mean; ratio versus \ac{CPF}, ${>}1.0$ indicates improvement). \ac{DARA} achieves the highest median throughput with the tightest interquartile range; \ac{OTIAS} achieves the lowest delay at the cost of substantially reduced throughput.}
\label{fig:ftp_boxplot}
\end{figure}

\begin{figure}[!htp]
\centering
\includegraphics[width=0.95\linewidth]{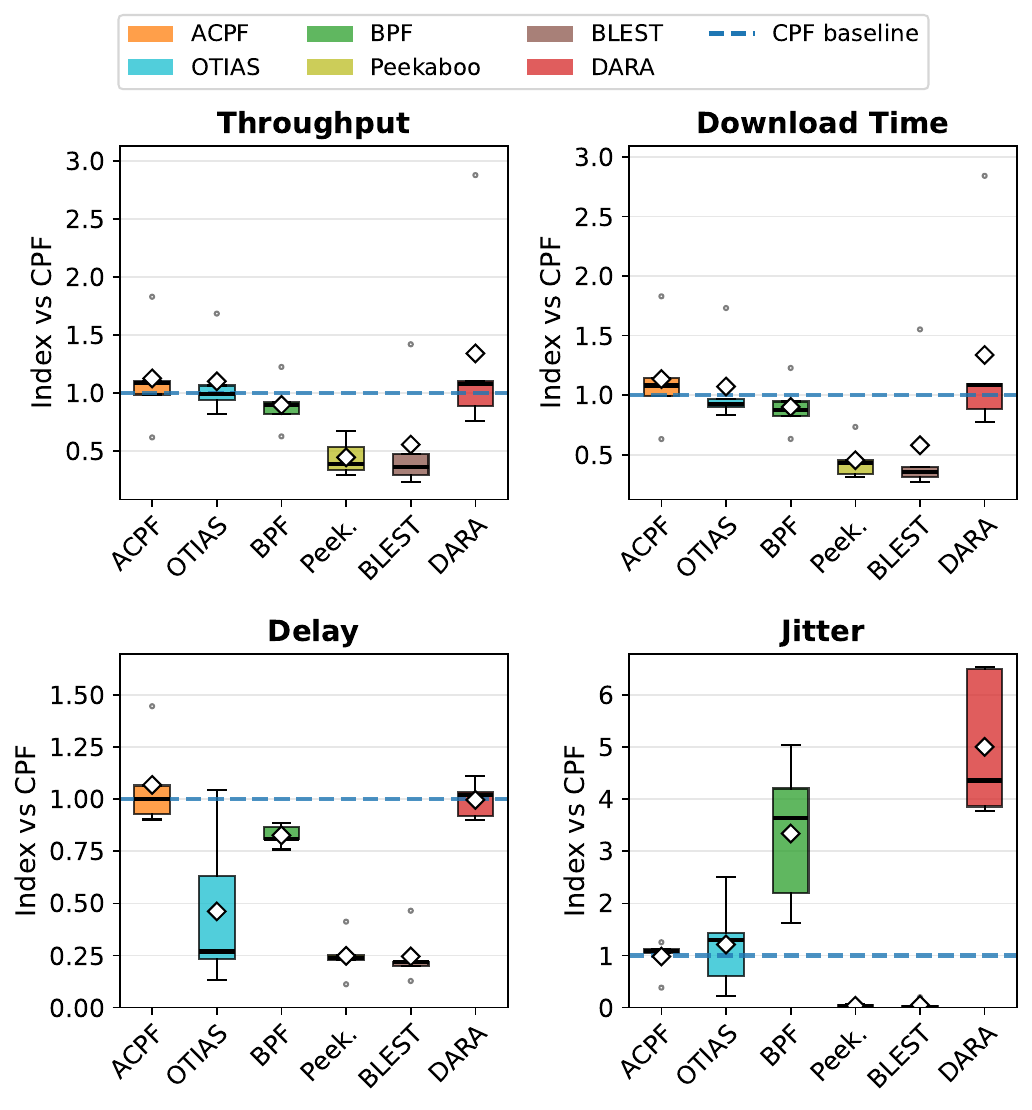}
\caption{QUIC file transfer performance distributions across five cellular traces (cURL application; diamond markers indicate the mean; ratio versus \ac{CPF}, ${>}1.0$ indicates improvement). Throughput gains are attenuated relative to TCP file transfer owing to nested congestion control interactions between tunnel-layer BBR and QUIC's internal algorithm. Peekaboo and \ac{BLEST} exhibit systematic collapse on delay and jitter axes, reflecting the combined effect of algorithmic limitations and \ac{MP-QUIC} per-stream ordering constraints rather than algorithmic quality alone.}
\label{fig:quic_boxplot}
\end{figure}

TCP file transfer is exercised via the FTP application protocol over
loss-based congestion control; QUIC file transfer is exercised via
cURL, where encrypted transport renders inner congestion control state
opaque to the scheduler, forcing allocation decisions based solely on
tunnel-layer metrics. Figure~\ref{fig:ftp_boxplot} presents FTP
performance distributions. \ac{DARA} achieves the largest throughput
gains on moderate-volatility traces, with performance peaking on
trace~4 before degrading on the extreme-volatility UMTS trace.
Crucially, \ac{DARA} remains functional on trace~5 whilst competitors
such as \ac{ACPF} and \ac{BPF} collapse to well below single-path
baseline. Figure~\ref{fig:quic_boxplot} presents QUIC performance,
where \ac{DARA} achieves the highest throughput on trace~1. Peekaboo
and \ac{BLEST} exhibit systematic delay and jitter collapse; \ac{MP-QUIC}'s
per-stream ordering constraints compound exploration-induced reordering,
so these results reflect the combined effect of scheduling algorithm
and transport protocol rather than algorithmic quality alone.


\subsubsection{Streaming Application Performance}

\begin{figure}[!htp]
\centering
\includegraphics[width=0.95\linewidth]{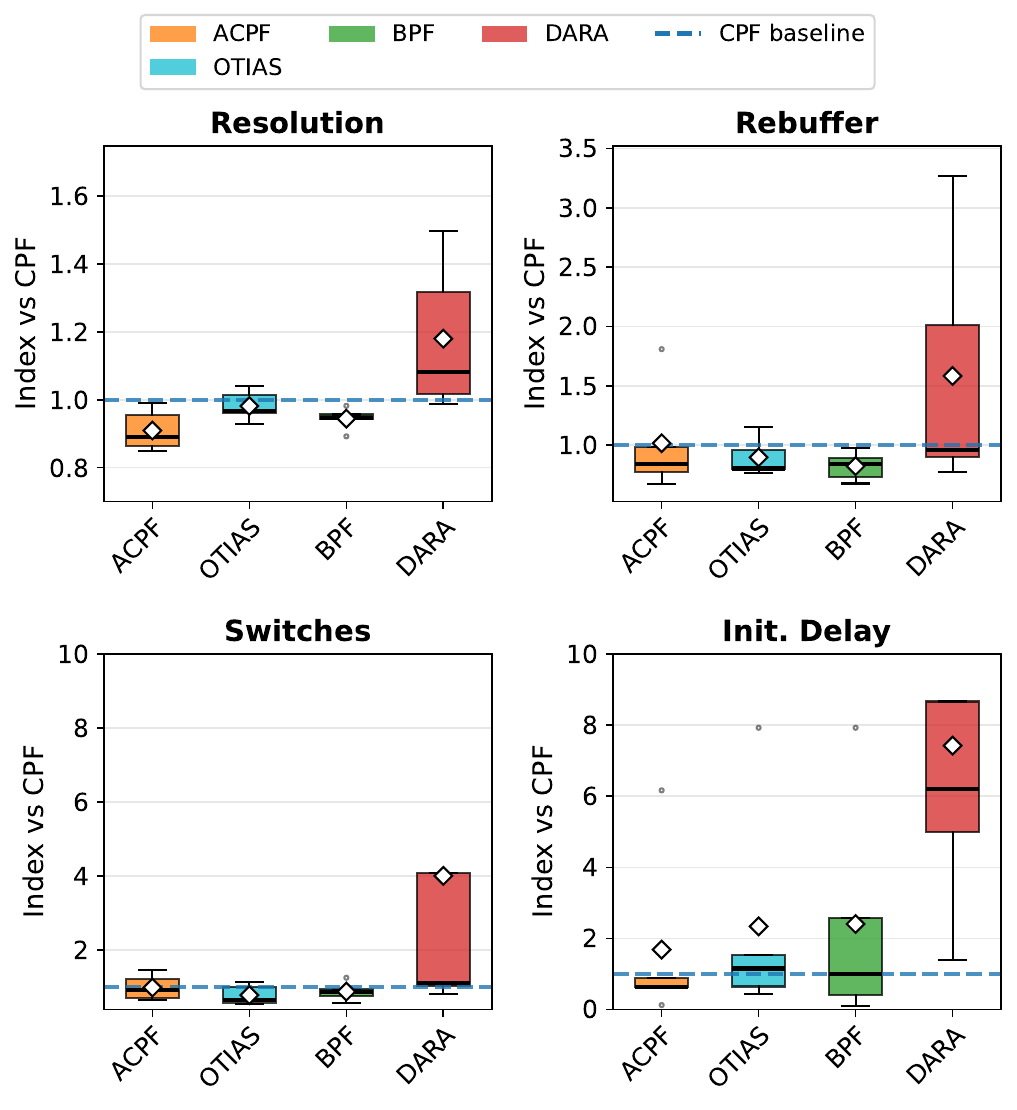}
\caption{YouTube \ac{QoE} distributions (diamond markers indicate
  the mean; ratio versus \ac{CPF}).}
\label{fig:youtube_boxplot}
\end{figure}

Figure~\ref{fig:youtube_boxplot} presents YouTube \ac{QoE}
distributions. \ac{DARA} achieves the highest resolution across
traces, with all competing schedulers remaining near or below the
\ac{CPF} baseline. The rebuffering index is above baseline for
\ac{DARA}, indicating more rebuffering events than \ac{CPF}, whilst
\ac{BPF} and \ac{OTIAS} achieve lower rebuffering through more
conservative allocation. The quality switch rate is markedly higher
for \ac{DARA}, consistent with more frequent adaptation to predicted
capacity changes rather than holding a fixed quality level. The
strongest and most consistent advantage is in initial delay, where
\ac{DARA} substantially outperforms all compared schedulers,
reflecting the benefit of preemptive capacity allocation at stream
startup. The overall picture is one of an aggressive scheduler that
prioritises peak quality and fast startup over stability, in contrast
to \ac{BPF} which sacrifices resolution for smoother playback.

\begin{figure}[!htp]
\centering
\includegraphics[width=0.95\linewidth]{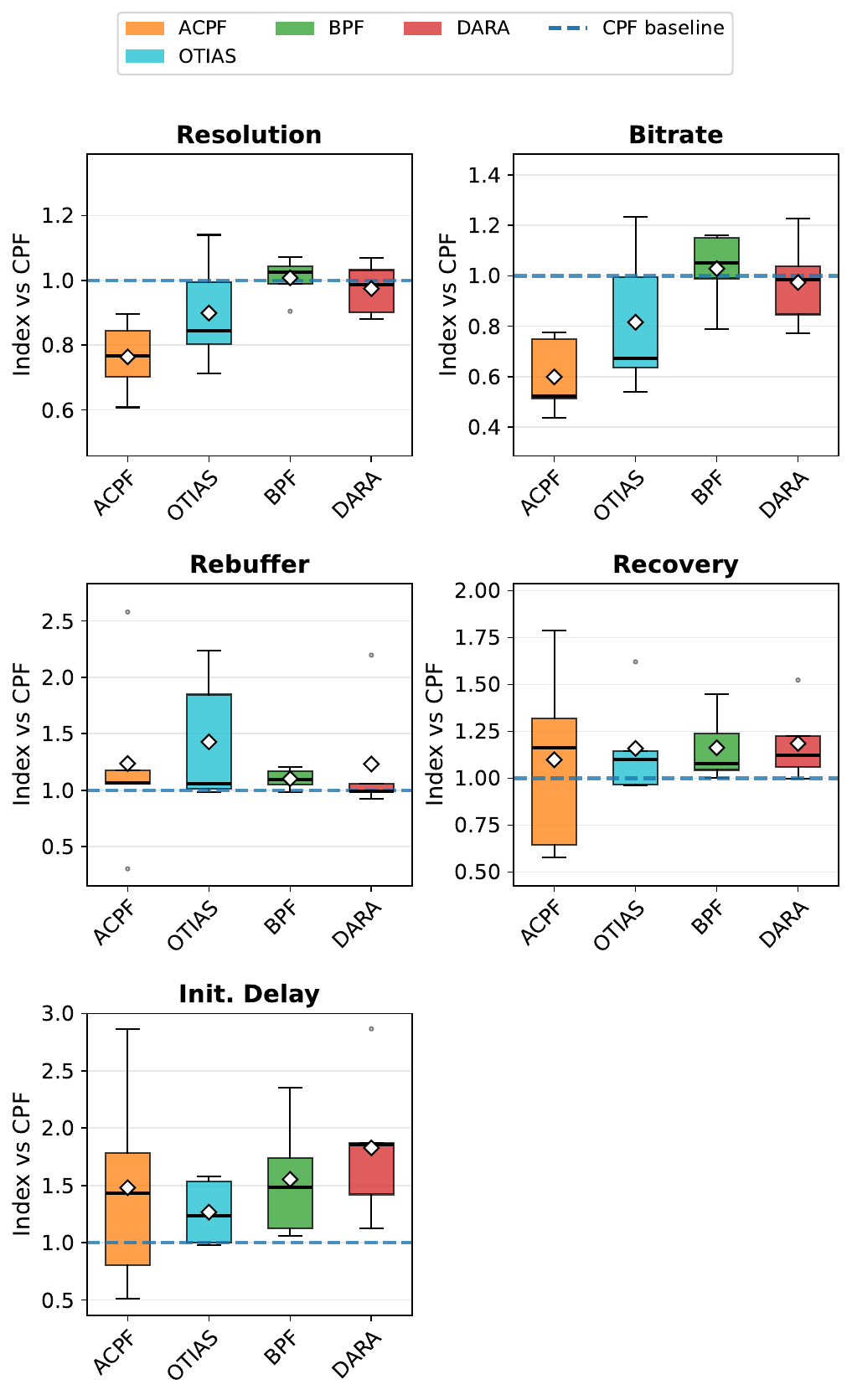}
\caption{Live streaming performance distributions (diamond markers
  indicate the mean; ratio versus \ac{CPF}).}
\label{fig:live_boxplot}
\end{figure}

\subsection{Summary}

The evaluation reveals systematic architectural constraints across
scheduler categories. Reactive schedulers exhibit observation-reaction
lag preventing effective burst exploitation, degrading as application
latency sensitivity increases. Coarse-grained prediction (\ac{BPF})
improves upon reactive approaches but fails under live streaming
constraints. State-of-the-art learning (Peekaboo, \ac{BLEST})
achieves competitive isolated throughput but with degraded latency
under high-volatility conditions.

\ac{DARA} achieves consistent leadership across applications:
\begin{itemize}
\item \textbf{File transfer:} Substantial throughput and jitter
  improvements on traces 1--4, with graceful degradation on the
  extreme-volatility UMTS trace whilst competitors collapse.
\item \textbf{YouTube:} Highest median resolution and substantially
  faster startup than all compared schedulers; higher switch rate
  and rebuffering index reflect more aggressive quality adaptation
  rather than conservative smoothing.
\item \textbf{Live streaming:} Significant initial delay reduction
  amplifying with volatility, and bounded rebuffering degradation
  relative to all compared schedulers.
\end{itemize}

\section{Conclusion}

This paper presents \ac{DARA}, an intelligent multipath scheduler for
high-mobility scenarios that predicts and exploits packet bursts,
reducing \ac{OFO} packets and \ac{HoL} blocking whilst improving
latency, jitter, throughput, and download times. Systematic ablation
studies confirm each component's contribution with large effect sizes
($d > 5$, bootstrap confidence intervals excluding zero) across all
baselines: predictor architecture selection confirms the Transformer's
advantage; depth analysis identifies the 3-layer configuration as
optimal; action-space analysis demonstrates the 5-level
discretisation outperforms coarser and finer alternatives; and reward
normalisation verification ensures optimised weights encode genuine
objective trade-offs.


\IEEEpubidadjcol

\addcontentsline{toc}{chapter}{Bibliography}
\bibliographystyle{ieeetr}
\bibliography{bibliography}
\vfill
\end{document}